\DeclareSymbolFont{AMSb}{U}{msb}{m}{n}
\documentclass[reqno,centertags,10pt,a4paper,noamsfonts]{amsart}
\pdfoutput=1
\usepackage[svgnames]{xcolor}
\usepackage{graphicx}
\usepackage[english]{babel}
\usepackage[babel=true,expansion=alltext,protrusion=alltext-nott,final]{microtype}
\usepackage[margin={3cm},marginparwidth={2.5cm}]{geometry}
\usepackage[utf8]{inputenc}
\usepackage{textcomp}
\usepackage[sb]{libertine}
\usepackage[libertine,bigdelims,vvarbb]{newtxmath}
\useosf
\usepackage[varqu,varl]{inconsolata}
\usepackage[supsfam=LibertinusSerif-Sup,%
       supscaled=1.2,%
       raised=-.13em]{superiors}
\usepackage{mathtools}
\usepackage[T1]{fontenc}
\usepackage{textcomp}
\usepackage{amsthm}
\usepackage[inline]{enumitem}
\numberwithin{equation}{section}
\usepackage[normalem]{ulem}

\usepackage[cal=cm]{mathalpha}
\usepackage{float}  
\usepackage{caption}
\usepackage{subcaption} 
\usepackage{xspace}         
\usepackage[scientific-notation=true]{siunitx}      
\usepackage{pstricks}
\usepackage{tikz}
\usetikzlibrary{positioning} 
\usetikzlibrary{calc}
\usepackage{pgfplots}
\pgfplotsset{width=10cm,compat=1.9}
\usepackage{bm}
\usepackage{sidecap}
\usepackage{graphicx,color}

\usepackage{amsmath}
\DeclareFontFamily{U}{mathx}{}
\DeclareFontShape{U}{mathx}{m}{n}{<-> mathx10}{}
\DeclareSymbolFont{mathx}{U}{mathx}{m}{n}
\DeclareMathAccent{\widehat}{0}{mathx}{"70}
\DeclareMathAccent{\widecheck}{0}{mathx}{"71}

\providecommand{\mr}[1]{\href{http://www.ams.org/mathscinet-getitem?mr=#1}{MR~#1}}
\providecommand{\zbl}[1]{\href{https://zbmath.org/?q=an:#1}{Zbl~#1}}

\providecommand{\doi}[2]{\href{https://doi.org/#1}{#2}}

\newcommand{\C}{\mathcal{C}}

\newcommand{\ii}{\mathrm{i}}

\definecolor{light_gray}{gray}{0.75}
\definecolor{lighter_gray}{gray}{0.5}
\colorlet{light_blue}{blue!20}
\definecolor{dark_green}{rgb}{0.0, 0.6, 0.0}
\definecolor{royal_blue}{rgb}{0.0, 0.22, 0.66}
\definecolor{salmon}{rgb}{1.0, 0.55, 0.41}
\definecolor{gold}{rgb}{0.8, 0.63, 0.21}
\definecolor{navy_blue}{rgb}{0.0, 0.0, 0.5}
\definecolor{crimson}{rgb}{0.79, 0.0, 0.09}
\definecolor{amethyst}{rgb}{0.6, 0.4, 0.8}
\definecolor{alizarin}{rgb}{0.82, 0.1, 0.26}
\definecolor{amaranth}{rgb}{0.9, 0.17, 0.31}
\definecolor{azure}{rgb}{0.0, 0.5, 1.0}
\definecolor{canaryyellow}{rgb}{0.82, 0.41, 0.12}
\definecolor{carrotorange}{rgb}{0.8, 0.33, 0.0}
\definecolor{cadmiumgreen}{rgb}{0.0, 0.42, 0.24}
\definecolor{copper}{rgb}{0.72, 0.45, 0.2}
\definecolor{aqua}{rgb}{0.5, 1.0, 0.83}
\definecolor{awesome}{rgb}{1.0, 0.13, 0.32}
\definecolor{candyapplered}{rgb}{1.0, 0.03, 0.0}
\definecolor{caribbeangreen}{rgb}{0.0, 0.8, 0.6}
\definecolor{indigo}{rgb}{0.0, 0.25, 0.42}

\DeclareMathOperator{\weaklystar}{\rightharpoonup\kern-2.2ex ^* \, \,}

\def\XXint#1#2#3{{\setbox0=\hbox{$#1{#2#3}{\int}$ }
\vcenter{\hbox{$#2#3$ }}\kern-.6\wd0}}

\newcommand{\R}{\mathbb R}
\newcommand{\N}{\mathbb N}
\newcommand{\Z}{\mathbb Z}
\renewcommand{\C}{\mathbb C}

\newcommand\norm[1]{\lVert #1 \rVert}

\newcommand\inner[1]{\langle #1 \rangle}

\newcommand\scpr{\boldsymbol{\cdot}}

\DeclareMathOperator{\Tr}{Tr}

\newcommand{\mL}{\mathrm{L}}

\renewcommand{\phi}{\varphi}

\newcommand{\mH}{\mathrm{H}}
\newcommand{\mW}{\mathrm{W}}

\newcommand{\ee}{\mathrm{e}}

\theoremstyle{plain}
\newtheorem{theorem}{Theorem}[section]
\newtheorem{proposition}[theorem]{Proposition}

\newtheorem{corollary}[theorem]{Corollary}
\newtheorem{lemma}[theorem]{Lemma}

\newtheorem*{theorem*}{Theorem}

\theoremstyle{definition}
\newtheorem{definition}[theorem]{Definition}
\newtheorem{remark}[theorem]{Remark}
\newtheorem*{remark*}{Remark}

\usepackage{xcolor}
\definecolor{JSTproofreadgreen}{RGB}{0,128,0}

\usepackage[pagebackref=false]{hyperref}
\hypersetup{
	linktoc=all,
	bookmarksdepth=3,
	colorlinks=true,
	linkcolor=Green,citecolor=Orange,urlcolor=DarkBlue,
	pdftitle={The unexpected regularity of DFT in 1D},
	pdfauthor={Thiago Carvalho Corso, Andre Laestadius}, 
	pdfsubject={},
	pdfkeywords={}} 
\begin{document}
\numberwithin{table}{section}

\title[A quantitative HK and the unexpected regularity of DFT]{A quantitative Hohenberg-Kohn theorem and the unexpected regularity of density functional theory in one spatial dimension}

\author[T.~Carvalho~Corso]{Thiago Carvalho Corso*}
\address[T.~Carvalho Corso]{Institute of Applied Analysis and Numerical Simulation, University of Stuttgart, Pfaffenwaldring 57, 70569 Stuttgart, Germany}
\email{thiago.carvalho-corso@mathematik.uni-stuttgart.de}

\author[A.~Laestadius]{Andre Laestadius}
\address[A.~Laestadius]{Department of Mathematical Sciences, Chalmers University of Technology and University of Gothenburg, SE-412 96 Gothenburg, Sweden}
\address[A.~Laestadius]{Hylleraas Centre for Quantum Molecular Sciences, Department of Chemistry, University of Oslo, 0315 Oslo, Norway}
\address[A.~Laestadius]{Department of Computer Science, Oslo Metropolitan University, 0130 Oslo, Norway}


\date{\today}
\thanks{\emph{Funding information}:  DFG -- Project-ID 442047500 -- SFB 1481. DFG -- Project-ID 572811220.
ERC -- Project-ID ~101041487.
RCN -- Project-ID 262695. \\[1ex]
\textcopyright 2025 by the authors. Faithful reproduction of this article, in its entirety, by any means is permitted for noncommercial purposes.}
\begin{abstract} In this paper we investigate the (Kohn--Sham) density-to-potential map in the case of spinless fermions in one spatial dimension, whose existence has been rigorously established by the first author in \cite{Cor26a}. Here, we focus on the regularity of this map as a function of the density and the coupling constant in front of the interaction term. More precisely, we first prove a quantitative version of the Hohenberg--Kohn theorem, thereby showing that this map is Lipschitz continuous with respect to the natural Sobolev norms in the space of densities and potentials. In particular, this implies that the inverse (Kohn--Sham) problem is not only well-posed but also Lipschitz stable. Using this result, we then show that the density-to-potential map is in fact real analytic with respect to both the density and the interaction strength. As a consequence, we obtain a holomorphic extension of the universal constrained-search functional to a suitable subset of complex-valued densities. This partially extends the DFT framework to non-self-adjoint Schr\"odinger operators. As further applications of these results, we also establish the existence of an exchange-only part of the exchange-correlation potential, and justify the G\"orling--Levy perturbation expansion for the correlation energy.
\end{abstract}
\setcounter{tocdepth}{1}
\maketitle
\setcounter{secnumdepth}{2}

\section{Introduction}

\subsection{Motivation} Across disciplines such as chemistry, materials science, and solid-state physics, density-functional theory (DFT) has become an indispensable tool for electronic-structure calculations \cite{KBP96,ED11,Bur12,Jon15,VT20,CF23}. By reformulating the many-body ground-state problem in terms of the single-particle density~\cite{HK64,Lev79,Lie83}, DFT provides a conceptually simpler framework for describing equilibrium states of many-body quantum systems. Its practical success, however, is largely due to the Kohn--Sham scheme \cite{KS65}, which introduces a fictitious non-interacting system that reproduces the same ground-state density as the interacting (physical) system of interest. Implicit in this framework is the notion of $v$-representability, i.e., that every interacting ground-state density corresponds to a ground state density of some non-interacting system. Furthermore, of central importance in the Kohn--Sham scheme is the exchange–correlation (xc) energy functional, whose differentiability properties are closely tied to the mathematical consistency of the entire framework. However, both the $v$-representability problem and the differentiability properties of the xc functional are rather subtle questions which are not yet fully understood \cite{THS+22,WAR+23}.
In fact, in the mathematical formulation presented by Lieb~\cite{Lie83}, it has been shown \cite{Lam07} that both the interacting and non-interacting universal functionals are everywhere discontinuous, therefore not differentiable. This raises the question of whether a differentiable formulation of (exact) DFT is even possible.

While these issues remain largely open in the physically most relevant setting of three-dimensional systems, significant progress has been made in the one-dimensional setting \cite{SPR+24,Cor25,Cor26b,Cor26a,SPR+25}. Indeed, by building on ideas of Sutter et al.\ \cite{SPR+24}, who considered a suitable class of distributional potentials, a complete characterization of the set of $v$-representable densities has been obtained by the first author \cite{Cor26a} in the one-dimensional setting. In particular, it has been shown that, for said class of external potentials, the set of $v$-representable densities is independent of the interaction potential and the xc-functional is (Gateaux) differentiable along this set. This shows that a differentiable formulation of DFT is indeed possible, thereby providing a rigorous mathematical framework for KS-DFT in this simplified setting. Moreover, with a differentiable framework in place, these results open the way to a systematic study of the regularity of the Kohn--Sham density-to-potential map. In particular, they naturally lead to the following questions:
\begin{enumerate}[label=(\roman*)]
\item How regular is the density-to-potential map? Put differently, if one varies the density, how does the potential change?
\item Furthermore, how does the representing potential change with respect to the interaction strength?
\end{enumerate}
These questions are not only natural from a mathematical perspective but also highly relevant in several applications of DFT.

The first question is crucial for the stability of the following inverse problem: given a target density, how can one reconstruct a potential that generates this density? When the target density is that of an interacting system and the potential is an effective one of a non-interacting system, this procedure is referred to as inverse Kohn--Sham (or Kohn--Sham inversion)~\cite{SW21}. In a more general way, one might rely on the Lieb maximization procedure to select a potential for a given ground-state density~\cite{TCH09}. While a characterization of $v$-representability and the Hohenberg--Kohn (HK) theorem settle, respectively, the existence and uniqueness of this inverse problem, the regularity of the density-to-potential map is intimately connected with its stability, i.e., whether a slight variation in the density yields a considerable change in the potential. Such stability can be observed, e.g., in the regularized inverse Kohn--Sham approach~\cite{PCL23,HBL25,PHH+25}, where the variations of the potential can be controlled by variations in the density and the inverse of the regularization parameter $\epsilon>0$. In the limit $\epsilon \downarrow 0$, however, the stability of the regularized scheme does not carry over to the exact setting,  as previously derived bounds blow-up in this limit (see~\cite{PL26} Corollary 6.3). In fact, it has been shown by Garrigue \cite{Gar21} that, within Lieb's DFT framework, the inverse problem is unstable (or ill-posed), due to compactness of the (ground-state) linear response operator.

The second question is related to the so-called (density-fixed) adiabatic connection (AC) \cite{PL75,GL76}, which plays a central role in both the theoretical investigation of DFT and the design of exchange--correlation approximations \cite{CF23}. More precisely, besides the implicitly assumed $v$-representability of the density (for a range of the interaction strength), these applications of the AC usually rely on an exact splitting of the xc-potential into exchange and correlation parts and on higher-order regularity of the adiabatic connection potential. This is the case, for instance, for the G\"orling--Levy  perturbation series for the correlation energy \cite{GL93,GL94}, which is employed in practical computations \cite{PDV24}. Nevertheless, to the best of our knowledge, neither an unambiguous splitting of the xc-potential nor the higher regularity (or even continuity) of the AC potential have been rigorously justified before.

In this paper, our goal is to address these questions in the simplified setting of spinless fermions living in one-dimensional space. Moreover, we emphasize that although our main concern is to demonstrate that several mathematical properties assumed to be true in standard (three-dimensional) DFT can actually be rigorously proven in the one-dimensional setting, this setting is also relevant in several applications, see~\cite{Gia04,Fog05,FP05,MMC+13,PKF19}.

\subsection{Main contributions} The main contributions of this paper can be summarized as follows:
\begin{enumerate}[label=(\roman*)]
    \item We prove a quantitative version of the Hohenberg--Kohn theorem, i.e., we show that, for a general but fixed interaction potential (or operator), the distance between external potentials can be bounded by the distance between the corresponding ground-state densities in their natural Sobolev spaces.
    \item We show that the constrained-search functional, and consequently the exchange--correlation functional, are real-analytic with respect to the density and the interaction strength. 
    \item As a surprising consequence of these results, we obtain a holomorphic extension of DFT to non-self-adjoint Schr\"odinger operators with complex-valued distributional potentials. More precisely, we show that the constrained-search functional and the density-to-potential map admit a holomorphic, respectively, biholomorphic extension to suitable subsets of complex-valued densities and potentials.
    
    \item We also present two further applications of these results. First, we establish the existence of an exchange-only potential, thereby providing a rigorous splitting of the xc-potential. Second, we justify the G\"orling--Levy perturbation series for the correlation energy.
\end{enumerate}
\subsection{Outline of the paper:} In the next section, we introduce the notation necessary to present our main results. The precise statements of these results and their applications are presented in Section~\ref{sec:main results}. We then sketch the main ideas for the proofs in Section~\ref{sec:proof sketch}. In Section~\ref{sec:discussion}, we briefly clarify how our results on the well-posedness of the inverse problem are related to, and do not contradict, the ill-posedness result by Garrigue \cite{Gar21}. Sections~\ref{sec:preliminaries},\ref{sec:QHK proof}, and~\ref{sec:analytic proof} are devoted to the proofs of our main results. More precisely, in Section~\ref{sec:preliminaries} we review some elementary results concerning the scale of Sobolev spaces induced by the (Neumann) Laplacian, which are used throughout our proofs. Section~\ref{sec:QHK proof} contains the detailed proof of the quantitative Hohenberg--Kohn theorem (cf. Theorem~\ref{thm:QHK}). The analyticity of the density-to-potential map (Theorem~\ref{thm:analytic map}) is proved in Section~\ref{sec:analytic proof}. The proofs of all applications discussed here are presented in Section~\ref{sec:corollaries proof}. In Section~\ref{sec:conclusion}, we conclude by summarizing our results and highlighting some natural questions that arise from these results. For the sake of completeness, we also recall the definition of (real) analytic and (complex) holomorphic functions in Banach spaces, and some of their fundamental properties in Appendix~\ref{app:analytic}.

\subsection*{Notation}

Let us now introduce the notation necessary to precisely state our main results. First, we denote by $I = (0,1)$ the open unit interval and by $I_N = (0,1)^N$ the open unit $N$-dimensional box. We set $\mH^1(I)$ for the first Sobolev space of $\mL^2$ integrable $\C$-valued functions with weak $\mL^2$ first derivative in $I$. Moreover, $\mH^1(I;\R) \subset \mH^1(I)$ will be used for the subspace of real-valued functions. We denote by $\mathcal{V}(I)$, or simply $\mathcal{V}$, the set of distributional potentials
\begin{align*}
    \mathcal{V}(I) \coloneqq  \mH^{-1}(I;\R),
\end{align*}
where $\mH^{-1}(I;\R)$ is the subspace of conjugate-linear continuous functionals in the Sobolev space $\mH^1(I)$ that are real-valued, i.e., $\inner{f,v} \in \R$ for any $f\in \mH^1(I;\R)$, where the brackets are used to denote the dual pairing between $v$ and $f$. Moreover, we let $\mathcal{V}/\{1\}$ be the quotient space of distributional potentials modulo additive constants
\begin{align*}
    \mathcal{V}/\{1\} \coloneqq \{ [v] : v,v' \in [v] \quad \mbox{if $v-v' = c$ for some constant $c\in \R$} \}
\end{align*}
endowed with the natural quotient norm
\begin{align}
    \norm{v}_{\mathcal{V}/\{1\}} \coloneqq  \inf_{\alpha \in \R} \norm{v-\alpha}_{\mH^{-1}}.\label{eq:quotient norm}
\end{align}
We also introduce the set of distributional interaction potentials
\begin{align}
    \mathcal{W}_p(I):=\mW^{-1,p}(I^2;\R),\qquad p>2,
  \qquad \mathcal{W}(I):=\bigcup_{p>2}\mathcal{W}_p(I),
\end{align} 
where $\mW^{-1,p}(I_2;\R)$ is the subspace of real-valued functionals in the anti-dual Sobolev space $\mW^{-1,p}(I_2) = (\mW^{1,\frac{p}{p-1}}(I_2))^\ast$. Any statement in which $w$ varies will be made
on a fixed $\mathcal{W}_p(I)$, never on the union. Nevertheless, unlike $\mathcal{V}(I)$, the choice of $\mathcal{W}(I)$ as the admissible set of interaction potentials is neither optimal nor essential for our results; it merely serves to illustrate how general the assumptions on the interaction potential can be. The reader can therefore think of the usual (translation-invariant) pointwise multiplication operators with potentials of the form $w(x,y) = w(\vert x-y\vert)$. In particular, this includes potentials like the Yukawa potential (in 1D as defined in \cite[Theorem 6.23]{LL01}), the soft-Coulomb and contact (delta) interactions.

For $N\in \N$, we denote by $\mathcal{H}_N$ the space of spinless (or spin-polarized) fermionic wave-functions, i.e., the $N$-fold anti-symmetric tensor product
\begin{align*}
    \mathcal{H}_N \coloneqq \wedge^N \mL^2(I).
\end{align*}
In addition, we define $\mathcal{H}_N^1$ as the subspace of fermionic wave-functions with finite kinetic energy, i.e.,
\begin{align*}
    \mathcal{H}_N^1 \coloneqq \mH^1(I_N) \cap \mathcal{H}_N
\end{align*}
endowed with the $\mH^1$ norm. We denote by $\mathcal{H}_N^{-1}$ the associated anti-dual space. In other words, $\mathcal{H}_N^{-1}$ is the subspace of continuous conjugate-linear functionals $\ell \in \mH^{-1}(I_N)$ that are anti-symmetric, i.e., satisfy
\begin{align*}
   ( S_N^\# \ell)(\Psi) = \ell(S_N^\# \Psi) = \ell(\Psi), \quad \mbox{for any $\Psi \in \mH^1(I_N)$,} 
\end{align*}
where $S_N^\#:\mH^1(I_N) \rightarrow \mathcal{H}_N^1$ is the anti-symmetrization operator
\begin{align*}
    (S_N^\# \Psi)(x) = \frac{1}{N!} \sum_{\sigma \in \mathcal{S}_N} \mathrm{sgn}(\sigma) \Psi(\sigma x), \quad x\in I_N,
\end{align*}
and $\mathcal{S}_N$ is the set of permutations of $N$ variables. 

For $v\in \mathcal{V}(I)$ and $w\in \mathcal{W}(I)$, we denote by $H_N(v,w)$ the self-adjoint operator
\begin{align*}
    H_N(v,w) = - \Delta + \sum_{i\neq j}^N w(x_i,x_j) + \sum_{j=1}^N v(x_j) \quad \mbox{with form domain $\mathcal{Q}_N = \mathcal{H}_N^1$.}
\end{align*}
More precisely, $H_N(v,w)$ is the unique self-adjoint operator associated to the sesquilinear form $\mathfrak{a}_{v,w}:\mathcal{H}_N^1\times \mathcal{H}_N^1 \rightarrow \C$ given by 
\begin{align}
    \mathfrak{a}_{v,w}(\Psi,\Phi) = \int_{I_N} \overline{\nabla \Psi(x)} \scpr \nabla \Phi(x) \mathrm{d}x + \inner{\rho_{\Psi,\Phi},v} + \inner{\rho^{(2)}_{\Psi,\Phi},w}, \label{eq:form def}
\end{align}
where $\rho_{\Psi,\Phi}$ and $\rho^{(2)}_{\Psi,\Phi}$ are respectively the mixed (or overlapping) single-particle and pair densities,
\begin{align*}
    &\rho_{\Psi,\Phi}(x) = N \int_{I_{N-1}} \Psi(x,x_2,...,x_N) \overline{\Phi(x,x_2,...,x_N)} \mathrm{d} x_2... \mathrm{d}x_N,\\
    &\rho_{\Psi,\Phi}^{(2)}(x,y) = N(N-1) \int_{I_{N-2}} \Psi(x,y,x_3,...,x_N)\overline{\Phi(x,y,x_3,...,x_N)}\mathrm{d} x_3... \mathrm{d} x_N.
\end{align*}
Here, $\inner{\rho, v}$ and $\inner{\rho^{(2)},w}$ stand for the dual pairings in the respective Sobolev spaces. This notation agrees with our convention for the inner product in $\mathcal{H}_N$, which is conjugate-linear in the left entry and linear in the right one. The fact that this pairing is well-defined and the quadratic form $\mathfrak{a}_{v,w}$ defines a semibounded self-adjoint operator $H_N(v,w)$ relies on the regularity of the reduced densities and the KLMN theorem, see, e.g., \cite[Section 3]{Cor26b}. To simplify the notation, we set $ \mathfrak{a}_{v,w}(\Psi) = \mathfrak{a}_{v,w}(\Psi,\Psi)$. 

\section{Main results}
\label{sec:main results}

We now turn to the main results of this paper and then discuss some of their applications. For this, we first recall that, for a fixed number of particles $N\in \N$ and a fixed interaction potential $w \in \mathcal{W}(I)$, the set of $\mathcal{V}$-representable densities is defined as 
\begin{align*}
    \mathcal{D}_N(w) \coloneqq \{ \rho_{\Psi} : \Psi \mbox{ is a (normalized) ground-state of $H_N(v,w)$ for some $v\in \mathcal{V}$}\}.
\end{align*}
According to the results in \cite{Cor26a}, the set of $\mathcal{V}$-representable densities is in fact independent of the interaction potential and given by 
\begin{align}
    \mathcal{D}_N(w) = \mathcal{D}_N \coloneqq \left\{ \rho \in \mH^1(I;\R) : \int_I \rho(x) \mathrm{d} x = N \quad \mbox{and}\quad \rho(x) >0\quad \mbox{for any $x\in [0,1]$}\right\}.
\end{align}
Here we shall always use the continuous representative of an $\mH^1(I)$ function.  In particular, we note that by compactness every $\rho\in\mathcal{D}_N$ satisfies
\begin{align*}
    \min\{ \rho(x) : x\in [0,1]\} =:a_\rho>0.
\end{align*}
Moreover, it was shown in \cite{Cor26a} that, for a fixed interaction potential, any density in this space is represented by a unique (modulo additive constants) potential $v\in \mathcal{V}$, i.e., the Hohenberg--Kohn (HK) theorem holds. In particular, this establishes the existence of a well-defined and bijective density-to-potential map
\begin{align*}
    \rho \in \mathcal{D}_N \mapsto v(\rho;w)\in \mathcal{V}/\{1\},
\end{align*}
where $\rho$ is the unique (by \cite[Theorem 2.1]{Cor26b}) ground-state density of $H_N(v(\rho;w),w)$.

Our first result here then shows that this map is Lipschitz continuous. 
\begin{theorem}[Quantitative Hohenberg--Kohn Theorem] \label{thm:QHK} Let $w\in \mathcal{W}(I)$ and $N\in \N$. Then the map $\rho \in \mathcal{D}_N \mapsto v(\rho;w) \in \mathcal{V}(I)/\{1\}$ is locally Lipschitz, i.e., for any $\rho \in \mathcal{D}_N$ there exist $\delta>0$ and $C = C(\rho,w)>0$ such that
\begin{align}
    \norm{v(\rho;w) - v(\rho';w)}_{\mathcal{V}/\{1\}} \leq C \norm{\rho-\rho'}_{\mH^1},
\end{align}
for any $\rho'\in \mathcal{D}_N$ with $\norm{\rho-\rho'}_{\mH^1} \leq \delta$.
\end{theorem}

Theorem~\ref{thm:QHK} can be seen as a quantitative version of the Hohenberg--Kohn theorem (QHK). Indeed, it quantifies the distance between representing potentials in terms of the distance between the represented densities. In particular, it shows that, if $\rho=\rho'$, then $v(\rho;w) = v(\rho';w)$ modulo additive constants, which is precisely the statement of the HK theorem. Moreover, this result shows that the inverse problem of retrieving the potential that generates a given density is not only well-posed in the Hadamard sense but also Lipschitz stable, for both the interacting and non-interacting cases.

The next result is about regularity properties of the density-to-potential map beyond the Lipschitz continuity stated above. To state it precisely, let us, for a fixed $w\in \mathcal W(I)$, set 
\begin{align}
    v(\rho;\lambda)\coloneqq v(\rho;\lambda w) \quad \mbox{for a real parameter $\lambda \in \R$.} \label{eq:density-to-potential map def}
\end{align}
Note that $v(\rho;\lambda)$ is related to the so-called adiabatic connection in DFT, i.e., $v(\rho;\lambda)$ is the unique potential such that the Hamiltonian $H_N(v(\rho;\lambda),\lambda w)$ has ground-state density $\rho$. The following result shows that this map is remarkably regular.
\begin{theorem}[Analytic density-to-potential map] \label{thm:analytic map} The map $(\rho,\lambda) \in \mathcal{D}_N \times \R \rightarrow v(\rho;\lambda)$ is real-analytic, i.e., belongs to $C^\omega(\mathcal{D}_N\times \R; \mathcal{V}/\{1\})$.
\end{theorem}

\begin{remark}[Tangent space of $\mathcal{D}_N$] \label{rem:tanget space} The set $\mathcal{D}_N$ is an open subset of the affine space 
\begin{align} 
\mathcal{X}_N \coloneqq \left\{\rho\in \mH^1(I;\R):\int_I\rho(x) \mathrm{d}x=N\right\}. \label{eq:tangentspace}
\end{align}
Therefore $\mathcal{X}_0 \coloneqq \{ f\in \mH^1(I;\R): \int f = 0\}$ can be naturally identified as the tangent space at any point $\rho \in \mathcal{D}_N$.
\end{remark}

\begin{remark}[Analytic maps]
By analytic, we mean that locally around any point $(\rho,\lambda)$ the map $v(\cdot;\cdot)$ can be written as an absolutely convergent series of multi-linear functionals on the tangent space $\mathcal{X}_0 \times \R$, i.e., 
\begin{align*}
    v(\rho + \sigma,\lambda +  \mu) = v(\rho;\lambda) + \sum_{n\geq 1} v^{(n)}_{(\rho,\lambda)} \left((\sigma,\mu),...,(\sigma,\mu)\right), \quad \mbox{(for small enough $(\sigma,\mu)$),}
\end{align*}
where $v^{(n)}_{(\rho,\lambda)}:(\mathcal{X}_0 \times \R)^n \rightarrow \mathcal{V}/\{1\}$ is $n$-linear and continuous.
For the precise statement, see Definition~\ref{def:analytic}. 
\end{remark}

\begin{remark}[Periodic case] For an odd number of electrons $N$, the same results can be proved in the periodic case, i.e., when $I = (0,1)$ is replaced by the torus $\mathbb{T}=\R /\Z$, or equivalently, the spaces of densities, potentials, and Sobolev wave-functions are replaced by their periodic counterparts. We conjecture that, up to degeneracies, the result still holds for an even number of electrons. However, in this case we cannot immediately apply the results of \cite{Cor26b,Cor26a} to carry out the proof.
\end{remark}

\subsection{Applications}
\label{sec:applications}
Let us now present some immediate corollaries of the previous results. For this, recall that the exchange--correlation (xc) functional is defined as
\begin{align} 
	\lambda E_{\rm xc}(\rho;\lambda) \coloneqq F_{\rm LL}(\rho;\lambda) - T_{\rm KS}(\rho) - \lambda E_H(\rho), \label{eq:xc}
\end{align}
where $F_{\rm LL}$ is the Levy--Lieb constrained-search functional
\begin{align}
F_{\rm LL}(\rho;\lambda) \coloneqq \min \{ \inner{\Psi, H_N(0,\lambda w) \Psi} : \Psi \in \mathcal{H}_N^1 , \quad \rho_\Psi = \rho\}, \label{eq:LL def}
\end{align}
$T_{\rm KS}$ is the Kohn--Sham kinetic energy functional
\begin{align*}
T_{\rm KS}(\rho) \coloneqq \min \left\{ \norm{\nabla \Psi}_{\mL^2(I_N)}^2 : \Psi \mbox{ Slater determinant with $\rho_\Psi = \rho$} \right\},
\end{align*}
and $E_H$ is the Hartree functional
\begin{align*}
E_{\rm H}(\rho) \coloneqq \inner{\rho\otimes \rho, w},
\end{align*}
with the corresponding Hartree potential $v_{\rm H}(\rho) = D_\rho E_{\rm H}(\rho) \in \mathcal{V}$ given by 
\begin{align}
    \inner{\delta, v_H(\rho)} = \inner{\delta \otimes \rho + \rho \otimes \delta, w}. \label{eq:Hartree potential}
\end{align}
\begin{remark}[Kohn--Sham kinetic energy] \label{rem:kinetic energy} In the particular setting studied here, the original Kohn--Sham kinetic energy defined in terms of Slater determinants agrees with the non-interacting Levy--Lieb functional on $\mathcal{D}_N$, i.e., $T_{\rm KS}(\rho) = F_{\rm LL}(\rho;0)$ for any $\rho \in \mathcal{D}_N$ (see \cite[Lemma 6.1]{Cor26a}). \end{remark}

Following \cite{LCP+24}, we define the exchange energy by the right derivative of $\lambda \mapsto F_{\rm LL}(\rho;\lambda)$ at $\lambda = 0$ minus the Hartree functional, 
\begin{align*}
E_x(\rho) \mathrel{\coloneqq} \lim_{\lambda \downarrow 0} \frac{ F_{\rm LL}(\rho;\lambda) - F_{\rm LL}(\rho;0) }{\lambda} - E_{\rm H}(\rho).
\end{align*}
This gives a rigorous definition\footnote{This definition of $E_{\rm x}$ also agrees with the coordinate-scaled, high-density limit definition of standard DFT in three dimensions.} of the exchange energy without reference to the function spaces of the densities in question. We define the correlation energy by
\begin{align*}
E_c(\rho;\lambda) \coloneqq E_{\rm xc}(\rho;\lambda)-E_x(\rho).
\end{align*}
Thus the xc-functional splits into exchange and correlation parts as
\begin{align*}
	\lambda E_{\rm xc}(\rho;\lambda) = \lambda E_x(\rho) + \lambda E_c(\rho;\lambda).
\end{align*}
The authors from \cite{LCP+24} then raised the question of whether a rigorous definition of an exchange-only potential is possible. Here we show that, in our setting, this is indeed the case. 

More precisely, in view of~\eqref{eq:xc} and Remark~\ref{rem:kinetic energy}, an immediate application of Theorem~\ref{thm:analytic map} shows that the exchange--correlation functional is actually analytic. This shows that the two-sided derivative at $\lambda = 0$ exists and agrees with the right derivative used to define $E_x$, and that $E_x$ is analytic in the density. Its density derivative therefore defines an exchange-only potential. To the best of our knowledge, this provides the first rigorous proof of an exact splitting of the xc-potential into exchange and correlation parts for continuum systems. We summarize these results in the next corollaries.
\begin{corollary}[Analytic Levy-Lieb functional] \label{cor:analytic levy-lieb} For any $N\in \N$ and $w\in \mathcal{W}(I)$, the Levy-Lieb functional $F_{\rm LL}$ defined in~\eqref{eq:LL def} belongs to $C^\omega(\mathcal{D}_N\times \R;\R)$.
\end{corollary} 

\begin{corollary}[Analytic exchange--correlation] \label{cor:exchange} For any $N\in \N$ and $w\in \mathcal{W}(I)$, the exchange--correlation energy defined in \eqref{eq:xc} belongs to $C^\omega(\mathcal{D}_N\times \R;\R)$. Therefore, the two-sided derivative exists and
\begin{align*}
E_x(\rho) = \partial_\lambda F_{\rm LL}(\rho;0) - E_{\rm H}(\rho), \quad \mbox{and} \quad E_x \in C^\omega(\mathcal{D}_N;\R).
\end{align*}
In particular, the exchange potential is well-defined in $\mathcal{V}/\{1\}$ by
\begin{align}
	v_{x}(\rho) \mathrel{\coloneqq} D_\rho E_{x}(\rho) \in \mathcal{V}/\{1\} , \quad \mbox{and it satisfies} \quad 
v_{x}(\rho) = -\partial_\lambda v(\rho;0) - v_{\rm H}(\rho) . \label{eq:exchange formula}
\end{align}
The correlation potential is consequently well-defined by
\begin{align*}
v_{\rm c}(\rho;\lambda) \coloneqq v_{\rm xc}(\rho;\lambda)-v_{\rm x}(\rho) = D_\rho E_{\rm c}(\rho;\lambda),
\end{align*}
where $v_{\rm xc}(\rho;\lambda) \coloneqq D_\rho E_{\rm xc}(\rho;\lambda)$.
\end{corollary}

Another immediate consequence of Theorem~\ref{thm:analytic map} is that the xc-functional admits a series expansion around $\lambda = 0$. In the physics literature, this expansion is known as the G\"orling--Levy perturbation series and was formally derived in \cite{GL93}. Here, we show that this expansion can be made rigorous, i.e., the series is absolutely convergent even at the level of potentials. 

\begin{corollary}[G\"orling--Levy perturbation expansion] \label{cor:GL perturbation series} For $\lambda \in \R$ small enough, the series
\begin{align*}
\lambda E_\mathrm{c}(\rho;\lambda) = \lambda^2 E^{\rm GL2}_\mathrm{c}(\rho) + \sum_{k \geq 3} E^{\rm GLk}_\mathrm{c}(\rho) \lambda^k,
\end{align*}
where $E^{\rm GL2}_c(\rho)$ is the second-order G\"orling--Levy perturbation term
\begin{align*}
    E^{\rm GL2}_c(\rho) = -\sum_{j\geq 1} \frac{|\inner{\Psi_0, \left(\hat{w} - \hat{v}_x(\rho) - \hat{v}_H(\rho)\right) \Psi_j}|^2}{\omega_j}
\end{align*}
and $E^{\rm GLk}_c$ are the higher-order terms, is absolutely convergent. Here, $v_H(\rho)$ is the Hartree potential~\eqref{eq:Hartree potential}, $\hat{v} = \sum_{k=1}^N v(x_k)$ and $\hat{w} = \sum_{i\neq k}^N w(x_i,x_k)$ are the $N$-particle operators associated to $v$ and $w$ (see Section~\ref{sec:sobolev scales}), and $\Psi_j$ and $\omega_j$ are respectively the $N$-particle eigenfunctions and excitation energies of the Kohn--Sham Hamiltonian $H_N(v(\rho;0),0)$.

Moreover, each term $E^{\rm GLk}_c(\rho)$ is analytic in $\mathcal{D}_N$ and the series
 \begin{align*}
     \lambda v_c(\rho;\lambda) = \sum_{k\geq 2} \lambda^k v^{\rm GLk}_c(\rho), \quad \mbox{where $v^{\rm GLk}_c(\rho) = D_\rho E^{\rm GLk}_c(\rho)$,}
 \end{align*}
 is absolutely convergent in $\mathcal{V}/\{1\}$ for $\lambda$ small enough.
\end{corollary}

\begin{remark}[Splitting of AC potential] The typical splitting of $v(\rho;\lambda)$ in the Kohn--Sham 
approach leads to
\begin{equation*}
 v_{\rm KS}(\rho):=   v(\rho;0) = v(\rho;\lambda) + \lambda v_{\rm H}(\rho) + \lambda v_{\rm xc}(\rho;\lambda),
\end{equation*}
where $v_{\rm xc} = D_\rho E_{\rm xc}$ and $v_{\rm H} = D_\rho E_{ \rm H}$. Note that here $v(\rho;\lambda)$ refers to the potential of the system with interactions modeled by $\lambda w$ having $\rho$ as the ground-state density. In applications, $v(\rho;1)$ is usually the potential of the physical system of fully interacting fermions. 
In our case, based on the above results, we can further split the exchange--correlation potential as
\begin{equation*}
    v_{\rm xc}(\rho;\lambda) = v_{\rm x}(\rho) + v_{\rm c}(\rho;\lambda),
\end{equation*}
where the exchange part is independent of $\lambda$.
\end{remark}

As a last application of Theorem~\ref{thm:analytic map}, we can extend the DFT framework to complex-valued potentials and complex-valued densities. To state this precisely, let us introduce the affine space of complex-valued densities,
\begin{align}
\mathcal{X}_N^\C\coloneqq \left\{\rho \in \mH^1(I) : \int_I \rho(x) \mathrm{d} x = N\right\}, \label{eq:complex-valued potentials}
\end{align}
and the corresponding space of complex-valued potentials modulo additive complex constants,
\begin{align}
    \mH^{-1}(I) /\{1\} \coloneqq \{ [v] : v,v'\in \mH^{-1}(I) ,\quad v \sim v' \quad \iff v-v' = \alpha \quad\mbox{for some $\alpha \in \C$} \} .
\end{align}
Then the following holds.

\begin{theorem}[Holomorphic extension of DFT] \label{thm:holomorphic DFT} Let $N\in \N$ and $w\in \mathcal{W}(I)$, then there exist a relatively open set $U \subset \mathcal{X}_N^\C \times \C$ and a holomorphic function $F_{\C}: U \rightarrow \C$ such that the following holds:
\begin{enumerate}[label=(\roman*)]
    \item(Extension) \label{it:extension} We have $\mathcal{D}_N \times \R \subset U$ and 
    \begin{align}
        F_\C(\rho;\lambda) = F_{\rm LL}(\rho;\lambda)\quad \mbox{for any $(\rho,\lambda) \in \mathcal{D}_N\times \R$,} \label{eq:extension property}
    \end{align}
    where $F_{\rm LL}$ is the Levy--Lieb functional defined in~\eqref{eq:LL def}.
    \item (Uniqueness) \label{it:uniqueness} The map $F_\C$ is unique in the following sense. For any other holomorphic map $F_\C':U' \rightarrow \C$ satisfying~\eqref{eq:extension property} we have $F_\C = F_\C'$ on the connected component of $U\cap U'$ containing $\mathcal{D}_N \times \R$.
    \item (Bijective density-to-potential map) \label{it:bijective potential map} The set
    \begin{align*}
        V \coloneqq \{(-D_\rho F_\C(\rho;\lambda), \lambda) : (\rho,\lambda) \in U\} \subset \mH^{-1}(I) /\{1\} \times \C,
    \end{align*}
    is open and the map $G: U \rightarrow V$ given by
    \begin{align*}
        (\rho,\lambda) \in U \mapsto G(\rho,\lambda) = (v_\C(\rho;\lambda), \lambda) \coloneqq (-D_\rho F_\C(\rho;\lambda), \lambda) \in V
    \end{align*}
    is a bi-holomorphic bijection.
    \item (Potential-to-density map via Hamiltonian) \label{it:Hamiltonian relation} There exists a rank-one projection-valued holomorphic map $P_\C : V \mapsto \mathcal{S}_1^1$ such that
    \begin{align*}
        G^{-1}(v,\lambda) = (\rho_{P_\C(v,\lambda)},\lambda), \quad \mbox{for $(v,\lambda) \in V$,}
    \end{align*}
    where $\rho_{P_\C(v,\lambda)}$ is the density associated to the rank-one projection $P_\C(v,\lambda)$. The space $\mathcal{S}^1_1$ is essentially the set of trace-class operators with finite kinetic energy, see Section~\ref{sec:trace spaces} for the precise definition. Moreover $P_\C$ satisfies
    \begin{align}
        \left[H_N(v,\lambda w) - \left(F_\C(G^{-1}(v,\lambda)) + \inner{\rho(v,\lambda), v}\right) \mathbb{1}\right] P_\C(v,\lambda) = 0, \label{eq:eigenvalue property}
    \end{align}
    for any $(v,\lambda) \in V$, where $\mathbb{1}$ is the identity operator in $\mathcal{H}_N$. Note that $P_\C(v,\lambda)$ is not necessarily self-adjoint (or equivalently an orthogonal projector) for complex $(v,\lambda)$.
\end{enumerate}
\end{theorem}

Let us now try to explain the results of Theorem~\ref{thm:holomorphic DFT} in a few words. The first property states that the constrained search functional $F_{\rm LL}$ has a holomorphic continuation to an open subset of complex densities and complex interaction strengths. Property~\ref{it:uniqueness} shows that such an extension is essentially unique. These are the standard extension and uniqueness statements for analytic functions, which still hold in infinite-dimensional Banach spaces (see Propositions~\ref{prop:identity theorem} and~\ref{prop:holomorphic extension} in Appendix~\ref{app:analytic}). 

Properties~\ref{it:bijective potential map} and~\ref{it:Hamiltonian relation} are more interesting and justify the name holomorphic DFT. The first one establishes the existence of a one-to-one mapping between complex densities and interaction strengths to complex potentials and interaction strengths. This provides an extension of the density-to-potential map in~\eqref{eq:density-to-potential map def}. Property~\ref{it:Hamiltonian relation} then shows that, like in the real-valued case, for any fixed interaction strength $\lambda \in \C$, any density in the sector $U \cap (\mH^1(I) \times \{\lambda \})$ of the domain of $F_\C$ is the density of a spectral Riesz projector of the Hamiltonian $H_N(v,w)$ with associated eigenvalue $F_\C(\rho;\lambda) + \inner{\rho,v}$, for any representative $v\in \mH^{-1}(I)$ in the equivalence class $-D_\rho F(\rho;\lambda) \in \mH^{-1}(I)/\{1\}$. Moreover, the equivalence class of representing potentials is unique by property~\ref{it:bijective potential map}, which can be understood as a complex counterpart of the HK theorem. 

However, in contrast with the real-valued case, the Hamiltonian is no longer self-adjoint. Consequently, $P_\C$ is not necessarily self-adjoint (thus not orthogonal) and the corresponding density not necessarily real-valued and positive. In fact, the density corresponds to the overlapping single-particle density of the left and right eigenfunctions of $H_N(v,w)$ and can attain any complex value. Moreover, we note that the meaning (or physical interpretation) of the associated eigenvalue 
\begin{align*}
    E_\C(v,\lambda) \coloneqq  F_\C(\rho_\C(v,\lambda),\lambda) + \inner{\rho_\C(v,\lambda),v}
\end{align*} is not clear, since there exists (to the best of our knowledge) no natural definition of a ground-state energy for non-self-adjoint Schr\"odinger operators. In fact, it is clear that for $(v,\lambda)$ close enough to the real sector $\mathcal{V}/\{1\} \times \R$, the energy $E_\C(v,\lambda)$ corresponds to the (unique) eigenvalue of $H_N(v,w)$ with the smallest real part. However, when $(v,\lambda)$ is far from the real sector, it is not clear (to us) whether this is still the case. In particular, we are not aware of a variational principle to evaluate the holomorphic constrained-search functional $F_\C(\rho;\lambda)$. 

Finally, let us emphasize that, at present, we are not aware of a direct physical application of this holomorphic extension; rather, we regard it as a mathematical consequence of the analyticity of DFT, whose potential significance remains to be explored.

\subsection{Proof strategy} \label{sec:proof sketch} Let us now outline the main steps in the proofs of Theorems~\ref{thm:QHK} and~\ref{thm:analytic map}. The proof of the quantitative Hohenberg--Kohn theorem relies on three main steps. In the first step, we obtain two estimates: a bound on the distance between the ground-state wave-functions generated by two potentials in terms of the difference between their densities times the distance between external potentials (see~\eqref{eq:H1est}), and a locally uniform bound on the representing potential (cf. Lemma~\ref{lem:unif bound potential}). The former estimate follows from an application of the variational principle and a standard energy-to-state estimate for general self-adjoint operators (see Lemma~\ref{lem:VP}). We emphasize, however, that this estimate relies on the non-degeneracy of the ground-state (or existence of a spectral gap), which is proved in \cite{Cor26b} for the systems considered here. The local uniform estimate on the representing potential relies on its connection with the Levy--Lieb functional and well-known upper and lower bounds for this functional in terms of the density. In the second step of the proof, we revisit the proof of the HK theorem but with potentials that generate different ground-state densities.  More precisely, we consider the difference Schr\"odinger equation $\inner{\Phi, \left(H_N(v,w) \Psi - H_N(v',w) \Psi'\right)}$ and apply the previous estimates to obtain a bound on $\inner{\rho_{\Psi,\Phi},v-v'}$ in terms of the difference between the densities and the potentials. The last and key step of the proof then consists in showing that the derivative of the wave-function-to-density map, i.e., the map 
\begin{align}
    \Phi \in \mathcal{H}_N^1\cap \{\Psi\}^\perp \mapsto B_\Psi(\Phi) \coloneqq \rho_{\Phi, \Psi} \in \mathcal{X}_0, \label{eq:derivative density map}
\end{align}
is, in fact, invertible from a suitable subspace of $\{\Psi\}^\perp \cap \mathcal{H}_N^1$ to the set of density variations $\mathcal{X}_0$ (see Lemma~\ref{lem:inverse}). This step crucially relies on (maximal) regularity properties of the reduced densities of general states, the strict positivity of the density, and the unique continuation property of the ground-state wave-function obtained in \cite{Cor26b}.

The main idea of the proof of Theorem~\ref{thm:analytic map} is to study the potential-to-density map and apply the implicit function theorem (IFT). The main advantage of studying the potential-to-density map instead of the density-to-potential map is that the former is rather explicit and can be shown to be analytic by standard perturbation theory arguments (i.e., resolvent expansions and contour integrals). The key difficulty in applying the IFT then consists of showing that the derivative of this map, namely the potential-to-density \emph{linear response operator} $D_v\rho$ (LRO), is invertible in the appropriate spaces. The injectivity of this operator is a consequence of the unique continuation of the ground-state wavefunction and the crucial Lemma~\ref{lem:inverse} establishing the surjectivity of the map $B_\Psi$ in~\eqref{eq:derivative density map}. The dense range property then follows from the fact that $D_v\rho$ is a symmetric operator from $\mathcal{V}/\{1\} \cong (\mathcal{X}_0)^\ast$ to $\mathcal{X}_0$, i.e., $(D_v \rho)^\ast = D_v\rho$. Finally, to show that the range of $D_v\rho$ is closed, and therefore $(D_v\rho)^{-1}$ is bounded, we rely on the quantitative HK result in Theorem~\ref{thm:QHK}. 

The proofs of Corollaries~\ref{cor:exchange} and~\ref{cor:GL perturbation series} are immediate applications of Theorem~\ref{thm:analytic map}. The proof of Theorem~\ref{thm:holomorphic DFT} relies on Theorem~\ref{thm:analytic map} and some elementary results on the theory of holomorphic functions in Banach spaces, which we collect in Appendix~\ref{app:analytic}.

\subsection{Connection with a  previous result on the instability of the inverse problem} \label{sec:discussion}

To conclude this section, let us briefly clarify the relation between our results and the work by Garrigue \cite{Gar21} on the ill-posedness of the inverse problem.

First, we recall that, in \cite{Gar21}, Garrigue shows that the inverse Kohn--Sham scheme, i.e., the problem of computing the effective potential whose corresponding non-interacting system generates a given ground-state density is ill-posed (in arbitrary dimensions). This is achieved by showing that the linear response operator $D_v\rho$ is compact, and therefore, the inverse map is unbounded. In particular, this leads to the conclusion that the inverse Kohn--Sham scheme is unstable.

In stark contrast, our results here show that the inverse Kohn--Sham scheme is very well-posed, i.e., the inverse map $\rho \mapsto v(\rho;w)$ is not only Lipschitz but also analytic. This apparent contradiction can be resolved by taking a closer look at the topologies considered for the space of potentials and densities. More precisely, in \cite{Gar21}, Garrigue considers $\mL^p$ and $\mW^{1,1}$ topologies respectively for the space of potentials and densities. While this choice is presumably motivated by Lieb's classical framework of DFT \cite{Lie83} (and the available unique continuation results for the specific range of exponents considered there \cite{Gar18}), such topologies are not well suited for the study of the (inverse) Kohn--Sham scheme (in the continuum case). Indeed, in our setting of one-dimensional systems, one can see that the compactness of the linear response operator in the $\mL^p$ spaces arises from the Sobolev embedding, rather than from any inherent lack of continuity of the inverse operator. More precisely, we note that, by the standard Sobolev embedding theorem in 1D, the space $\mL^1(I;\R)$ is compactly embedded in $\mathcal{V} = \mH^{-1}(I;\R)$. Therefore, the LRO $D_v \rho : \mathcal{V} \rightarrow \mathcal{X}_0$ becomes compact when restricted to an operator from $\mL^1(I)$ to $\mathcal{X}_0$ as investigated in \cite{Gar21}. Similarly, this operator becomes compact if we enlarge the co-domain to any space into which $\mathcal{X}_0$ is compactly embedded, e.g., $\mL^\infty(I)$. These observations highlight the importance of considering appropriate topologies for the space of densities and potentials when studying the inverse problem.

\section{Mathematical preliminaries} \label{sec:preliminaries}

In this section, we set up some additional notation and recall some elementary results regarding the $N$-particle Sobolev spaces $\mathcal{H}_N^{\pm 1}$ and their reduced densities.

\subsection{Operators and resolvents on the Sobolev scales} \label{sec:sobolev scales} Throughout the paper, we denote by $\mathcal{B}_{\pm 1,\pm 1} = \mathcal{B}(\mathcal{H}_N^{\pm 1}, \mathcal{H}_N^{\pm 1})$ the spaces of bounded linear operators from $\mathcal{H}_N^{\pm 1}$ to $\mathcal{H}^{\pm 1}_N$ endowed with the respective operator norms. Similarly, we use, respectively, the abbreviations $\mathcal{B}_{\pm 1, 0}$, $\mathcal{B}_{0,\pm 1}$, and $\mathcal{B}_{0,0}$ for the spaces of bounded operators $\mathcal{B}(\mathcal{H}_N^{\pm 1}, \mathcal{H}_N)$, $\mathcal{B}(\mathcal{H}_N,\mathcal{H}_N^{\pm 1})$, and $\mathcal{B}(\mathcal{H}_N,\mathcal{H}_N) = \mathcal{B}(\mathcal{H}_N)$.

Recall that any $\Psi \in \mathcal{H}_N$ can be identified with an element in the anti-dual $\mathcal{H}_N^\ast$ via the (Riesz-Fréchet) mapping
\begin{align}
    \inner{\Phi,\mathcal{R}_\Psi} \coloneqq \inner{\Phi, \Psi}_{\mathcal{H}_N}. \label{eq:Riesz map}
\end{align}
Consequently, we have the sequence of continuous $\C$-linear inclusions
\begin{align*}
    \mathcal{H}_N^1 \hookrightarrow \mathcal{H}_N \overset{\mathcal{R}}{\hookrightarrow} \mathcal{H}_N^\ast \hookrightarrow \mathcal{H}_N^{-1}.
\end{align*}
Here, $\mathcal{H}_N^{-1}$ is chosen as the \emph{anti-dual} of $\mathcal{H}_N^1$ precisely to make the above inclusions $\C$-linear. Moreover, these inclusions imply that
\begin{align}
 \mathcal{B}_{s,t}\hookrightarrow\mathcal{B}_{s',t'}
  \quad\text{whenever}\quad s\leq s'\ \text{and}\ t\geq t', \quad s,s',t,t'\in \{-1,0,1\}.\label{eq:operator inclusions}
\end{align}

Let us now briefly clarify our convention for adjoints on the different operator spaces defined above. First, for any $A\in \mathcal{B}_{0,0}$, we define the adjoint $A^\ast \in \mathcal{B}_{0,0}$ in the usual way, i.e., via the identity
\begin{align*}
    \inner{\Psi, A \Phi}_{\mathcal{H}_N} = \inner{A^\ast \Psi, \Phi}_{\mathcal{H}_N}, \quad \mbox{for all $\Psi, \Phi \in \mathcal{H}_N$.}
\end{align*}
For $A \in \mathcal{B}_{-1,1}$, we can similarly define the adjoint $A^\ast$  as the unique operator in the same space, $\mathcal{B}_{-1,1}$, such that
\begin{align}
    \inner{AF, G} = \overline{\inner{A^\ast G,F}} \quad \mbox{for any $F,G\in \mathcal{H}_N^{-1}$,} \label{eq:adjoint relation}
\end{align}
where the overline denotes complex conjugation. The existence and uniqueness of the adjoint is guaranteed since $\mathcal{H}_N^{-1}$ is a Hilbert space (hence reflexive). Likewise, we define the adjoint of an operator $A\in \mathcal{B}_{1,-1}$ as the unique element $A^\ast \in \mathcal{B}_{1,-1}$ such that
\begin{align*}
    \inner{\Phi,A\Psi} = \overline{\inner{\Psi,A^\ast \Phi}}, \quad \mbox{for any $\Psi, \Phi \in \mathcal{H}_N^1$.}
\end{align*}
Again, existence, uniqueness, and boundedness of the adjoint are guaranteed. In fact, it follows that $\norm{A} = \norm{A^\ast}$ by duality. In addition, note that taking adjoints commutes with the inclusions $\mathcal{B}_{-1,1} \hookrightarrow  \mathcal{B}_{0,0} \hookrightarrow \mathcal{B}_{1,-1}$ mentioned in~\eqref{eq:operator inclusions}.

We now discuss the relation between $\mL^2$-orthogonal projectors and resolvents with the above scale of spaces. First, we note that, for any normalized $\Psi \in \mathcal{H}_N^1$, the $\mL^2$-orthogonal projector, defined as
\begin{align}
    P_\Psi (\Phi) = \inner{\Psi, \Phi}_{\mathcal{H}_N} \Psi, \quad \mbox{for any $\Phi \in \mathcal{H}_N$,} \label{eq:projector def}
\end{align}
can be extended to an operator in $\mathcal{B}_{-1,1}$ whose kernel is precisely the annihilator set
\begin{align}
    \ker P_\Psi = \mathcal{H}_N^{-1} \cap \{\Psi\}^\perp \coloneqq \{ F \in \mathcal{H}_N^{-1} : \inner{\Psi, F} = 0\}. \label{eq:annihilator def}
\end{align}
Indeed, this simply follows by using the extended formula
\begin{align}
    (P_\Psi F) = \inner{\Psi, F} \Psi, \quad \mbox{for any $F\in \mathcal{H}_N^{-1}$,} \label{eq:projector extension}
\end{align}
which agrees with~\eqref{eq:projector def} via the Riesz identification~\eqref{eq:Riesz map}. For later reference, let us summarize this observation in the next proposition. 

\begin{proposition}[$\mL^2$-orthogonal projectors on Sobolev scale] Let $\Psi \in \mathcal{H}_N^1$ be normalized, then the projector $P_\Psi$ in~\eqref{eq:projector def}
admits an extension to $\mathcal{B}_{-1,1}$ given by~\eqref{eq:projector extension}. In particular, $P_\Psi$ belongs to all of the operator spaces $\mathcal{B}_{s,t}$ with $s,t\in \{-1,0,1\}$. Consequently, the complementary projection $P_\Psi^\perp = 1-P_\Psi$ belongs to $\mathcal{B}_{-1,-1}$, $\mathcal{B}_{0,0}$ and $\mathcal{B}_{1,1}$. Moreover, $P_\Psi$  (hence also $P_\Psi^\perp$) is self-adjoint in the sense that $P_\Psi^\ast = P_\Psi$, where the adjoint is defined via~\eqref{eq:adjoint relation}.
\end{proposition}

Another useful consequence of the anti-dual convention is the following. Recalling that the forms in~\eqref{eq:form def} are Laplace bounded (see \cite[Section 3]{Cor26b}), the Hamiltonian $H_N(v,w)$ can be seen as a linear bounded operator in $\mathcal{B}_{1,-1}$. Precisely, $H_N(v,w) \in \mathcal{B}_{1,-1}$ acts as
\begin{align*}
    \inner{\Phi,H_N(v,w) \Psi} = \mathfrak{a}_{v,w}(\Phi,\Psi) \quad \mbox{with $\mathfrak{a}_{v,w}$ defined in~\eqref{eq:form def}.} 
\end{align*}
In particular, for any $z$ in the resolvent set of $H_N(v,w)$, the resolvent operator
\begin{align*} R_z(v,w) \coloneqq \left(z- H_N(v,w)\right)^{-1} \end{align*}
admits an extension to $\mathcal{B}_{-1,1}$ given by the inverse of $z-H_N(v,w)$ as an operator in $\mathcal{B}_{1,-1}$. This extension can be given explicitly in terms of the extended spectral projections in~\eqref{eq:projector extension} as
\begin{align}
    R_z(v,w) F = \sum_{j=0}^\infty \frac{1}{z-E_j} P_j(F), \quad \mbox{for any $F \in \mathcal{H}_N^{-1}$,} \label{eq:resolvent representation}
\end{align}
where $E_j$ denotes the $j^{th}$ eigenvalue of $H_N(v,w)$ and $P_j$ the associated spectral projector. Recall that $H_N(v,w)$ has discrete spectrum\footnote{The representation in~\eqref{eq:resolvent representation} still holds in the case of continuous spectrum by replacing sum by integral.} (as its form domain is compactly embedded in $\mathcal{H}_N$) so that $P_j$ is a sum of finitely many rank-one projectors of the form~\eqref{eq:projector def}. 

Similarly, the operator $E_0 - H_N(v,w)$, where $E_0$ is the ground-state energy of $H_N(v,w)$, is an isomorphism from the space $\mathcal{H}_N^1 \cap \{\Psi\}^\perp$ to the annihilator space $\mathcal{H}_N^{-1} \cap \{\Psi\}^\perp$ defined in~\eqref{eq:annihilator def}, where $\Psi$ denotes the ground-state wave-function of $H_N(v,w)$. Its inverse is given by the restriction of the reduced resolvent
\begin{align}
    R^\perp(v,w) = \lim_{z \rightarrow E_0} P_{\Psi}^\perp R_z(v,w) P_\Psi^\perp = \sum_{j\geq 1} \frac{1}{-\omega_j} P_j \label{eq:reduced resolvent}
\end{align}
to the annihilator space $\mathcal{H}^{-1}_N \cap \{\Psi\}^\perp$, where $\omega_j = E_j - E_0$ are the excitation energies (or frequencies) of $H_N(v,w)$. Let us also turn these observations into a proposition for later reference.

\begin{proposition}[Reduced resolvent] Let $R^\perp(v,w)$ be the reduced resolvent defined in~\eqref{eq:reduced resolvent}. Then $R^\perp(v,w) :\mathcal{H}_N^{-1}\cap \{\Psi\}^\perp \rightarrow \mathcal{H}_N^1 \cap \{\Psi\}^\perp$ is an isomorphism, i.e., is bounded and has a bounded inverse. Moreover, the following holds:
\begin{enumerate}[label=(\roman*)]
\item(Symmetry) \label{it:symmetry} The reduced resolvent is self-adjoint $(R^\perp)^\ast = R^\perp$ in the sense that
\begin{align}
   \inner{R^\perp(v,w) F,G} = \overline{\inner{R^\perp(v,w)G,F}}, \quad\mbox{for any $F,G \in \mathcal{H}_N^{-1}$,} \label{eq:adjoint resolvent}
\end{align}
\item(Strictly negative) \label{it:strictly negative} There exists $c>0$ such that
\begin{align}
    -\inner{R^\perp(v,w) F,F} \geq c\norm{P_\Psi^\perp F}_{\mathcal{H}_N^{-1}}^2, \quad \mbox{for any $F \in \mathcal{H}_N^{-1}$.} \label{eq:strictly negative}
\end{align}
\end{enumerate}
\end{proposition}
\begin{proof} Recalling that the forms induced by $v$ and $w$ are infinitesimally Laplace form-bounded (see \cite[Appendix A]{Cor26a}), we have the equivalence of norms\footnote{The equivalence of norms follows from the KLMN-type estimate used to show that the quadratic form $\mathfrak{a}_{v,w}$ is closed.}
\begin{align}
    \norm{\Psi}_{\mathcal{H}^1_N}^2 \lesssim \norm{\Psi}_{\mathfrak{a}_{v,w}} \coloneqq \mathfrak{a}_{v,w}(\Psi) + (1-E_0) \norm{\Psi}^2 \lesssim \norm{\Psi}_{\mathcal{H}_N^1}^2. \label{eq:norm equivalence}
\end{align}
Moreover, recalling that the spectral gap $\omega_1 = E_1 - E_0>0$ is positive by the non-degeneracy theorem in \cite{Cor26b}, we have
\begin{align}
    \inner{\Phi,(H_N(v,w) - E_0)\Phi}
    =\mathfrak a_{v,w}(\Phi)-E_0\norm{\Phi}_{\mathcal H_N}^2
    \geq \omega_1\norm{P_\Psi^\perp \Phi}_{\mathcal H_N}^2, \quad \Phi \in \mathcal{H}_N^1. \label{eq:some coercivity}
\end{align}
Together, estimates~\eqref{eq:norm equivalence} and~\eqref{eq:some coercivity} imply that, for some $\gamma>0$,
\begin{align}
    \inner{\Phi,(H_N(v,w) - E_0)\Phi}\geq
    \gamma\norm{\Phi}_{\mathcal H_N^1}^2,
    \qquad \Phi\in \mathcal{H}_N^1 \cap \{\Psi\}^\perp.\label{eq:coercive est}
\end{align}
Hence, by the usual Lax--Milgram argument/theorem, $H_N(v,w) - E_0$ is an isomorphism from $\mathcal{H}_N^1 \cap \{\Psi\}^\perp$
to $\mathcal H_N^{-1}\cap\{\Psi\}^\perp$ and the inverse is precisely $-R^\perp(v,w)$. 

The proof of~\ref{it:symmetry} is immediate. To prove~\ref{it:strictly negative}, let $F\in \mathcal{H}_N^{-1} \cap \{\Psi\}^\perp$ and set $\Phi = -R^\perp(v,w) F$. Then by~\eqref{eq:coercive est}
\begin{align*}
    -\inner{R^\perp(v,w)F,F}
    =\inner{\Phi,(H_N(v,w) - E_0)\Phi}
    \geq \gamma\norm{\Phi}_{\mathcal H_N^1}^2.
\end{align*}
On the other hand,
\begin{align*}
    \norm{F}_{\mathcal H_N^{-1}}
    =\norm{(H_N(v,w) - E_0)\Phi}_{\mathcal H_N^{-1}}
    \leq \norm{(H_N(v,w) - E_0)}_{\mathcal B_{1,-1}}
           \norm{\Phi}_{\mathcal H_N^1}.
\end{align*}
Together these estimates prove~\eqref{eq:strictly negative}.
\end{proof}

Similarly, we can associate to any external potential $v \in \mathcal{V}$ and any interaction potential $w\in \mathcal{W}(I)$, the $N$-particle one-body and two-body operators $\hat{v} \in \mathcal{B}_{1,-1}$ and $\hat{w} \in \mathcal{B}_{1,-1}$ via 
\begin{align}
    \inner{\Phi,\hat{v} \Psi} = \inner{\rho_{\Phi,\Psi},v} \quad \mbox{respectively} \quad \inner{\Phi,\hat{w} \Psi} = \inner{\rho^{(2)}_{\Phi,\Psi},w}. \label{eq:quantized operators}
\end{align}
These are the usual second quantization versions of $v$ and $w$ restricted to the sector of $N$ particles, which are sometimes denoted as $\hat{v} = \mathrm{d} \Gamma(v)$ in the literature. As they will be frequently used in the next sections, we collect some of their properties in the next proposition.

\begin{proposition}[Quantized operators]
The map $v \rightarrow \hat{v}$ is an injective and bounded operator from $\mH^{-1}(I;\C)$ to $\mathcal{B}_{1,-1}$, from $\mL^2(I)$ to $\mathcal{B}_{1,0}$, and from $\mH^1(I)$ to $\mathcal{B}_{1,1}$. Moreover, the adjoint operator $\hat{v}^\ast$ is given by
\begin{align}
    \hat{v}^\ast  = \hat{\overline{v}} \quad \mbox{where $\overline{v}$ is the complex conjugated functional} \quad \inner{\rho,\overline{v}}= \overline{\inner{\overline{\rho},v}}, \quad \mbox{for $\rho \in \mH^1(I)$.} \label{eq:adjoint potentials}
\end{align}
\end{proposition}

\subsection{Trace class operators in the Sobolev scale} \label{sec:trace spaces} We now introduce a scale of trace-class operators in the Sobolev spaces. First, we denote by $\mathcal{S}_1(\mathcal{H}_N)$ or simply $\mathcal{S}_1$ the standard space of trace-class operators in $\mathcal{H}_N$, i.e., the set of compact operators $A \in \mathcal{B}_{0,0}$ with finite trace norm
\begin{align*}
    \norm{A}_{\mathcal{S}_1} = \mathrm{Tr} \, |A|<\infty.
\end{align*}

Next, let us define $\mathcal{C}_\Delta : \mathcal{B}_{-1,1} \rightarrow \mathcal{B}_{0,0}$ as the conjugation by $(-\Delta+1)^{\frac12}$, i.e., 
\begin{align*}
    \mathcal{C}_\Delta A = (-\Delta +1)^{\frac12} A (-\Delta+1)^{\frac12}, \quad A \in \mathcal{B}_{-1,1},
\end{align*}
where $-\Delta$ denotes the Neumann Laplacian on $\mathcal{H}_N$. Note that, since $(-\Delta +1)^{\frac12}$ is an isomorphism\footnote{This follows from the same arguments used to show that $-\Delta+1$ is an isomorphism in $\mathcal{B}_{1,-1}$, namely, an application of the spectral theorem.} in both $\mathcal{B}_{0,-1}$ and $\mathcal{B}_{1,0}$, the conjugation map $\mathcal{C}_\Delta : \mathcal{B}_{-1,1} \rightarrow \mathcal{B}_{0,0}$ is also an isomorphism between operator spaces. The inverse $C_\Delta^{-1}: \mathcal{B}_{0,0} \rightarrow \mathcal{B}_{-1,1}$ is given by the inverse conjugation
\begin{align*}
           C_\Delta^{-1}(A) = (-\Delta+1)^{-\frac12} A (-\Delta + 1)^{-\frac12}.
\end{align*}

We can thus introduce the following space of trace-class operators with finite kinetic energy
\begin{align}
    \mathcal{S}_1^1 \coloneqq \{A \in \mathcal{B}_{-1,1}: \mathcal{C}_\Delta A \in \mathcal{S}_1 \}, \label{eq:kinetic trace space}
\end{align}
endowed with the norm
\begin{align*}
    \norm{A}_{\mathcal{S}^1_1} = \norm{\mathcal{C}_\Delta (A)}_{\mathcal{S}^1} = \mathrm{Tr} |(-\Delta+1)^{\frac12} A (-\Delta +1)^{\frac12}|.
\end{align*}
Since $\mathcal{S}_1$ is a Banach space and $\mathcal{C}_\Delta$ is an isomorphism, the space $\mathcal{S}_1^1$ is also a Banach space. Moreover, we note that this space is a standard construction, see e.g. \cite{BDF76}. 

We now recall the well-known duality between trace-class operators and bounded operators. Precisely, any element $F \in \mathcal{S}_1^\ast$ can be uniquely identified with an element $B_F \in \mathcal{B}_{0,0}$ via the duality relation
\begin{align*}
    \inner{A,F} = \mathrm{Tr}\,  A^\ast B_F ,
\end{align*}
where $A^\ast \in \mathcal{B}_{0,0}$ is the usual Hilbert space adjoint. Using this result, we can identify the anti-dual $(S_1^1)^\ast$ with the space $\mathcal{B}_{1,-1}$ as follows. For any $F \in (S_1^1)^\ast$ there exists a unique $B_F \in \mathcal{B}_{1,-1}$ such that
\begin{align*}
    \inner{A,F} = \mathrm{Tr} \,  (\mathcal{C}_\Delta A)^\ast (\mathcal{C}_\Delta^{-1}B_F) = \mathrm{Tr} \, A^\ast B_F.
\end{align*}

In a similar way, one could define $\mathcal{S}_1^{-1} \subset \mathcal{B}_{1,-1}$ as the pull-back of the trace-class operators via the inverse conjugation $\mathcal{C}_\Delta^{-1} : \mathcal{B}_{1,-1} \rightarrow \mathcal{B}_{0,0}$, and identify the dual $(\mathcal{S}_1^{-1})^\ast$ with $\mathcal{B}_{-1,1}$. However, these spaces will not be used here.

\subsection{The density map} \label{sec:density map} We now introduce and discuss a few elementary properties of the density map. First, the density map can be defined for rank-one operators of the form $P_{\Psi,\Phi}(\chi) = \Psi \inner{\Phi, \chi}_{\mathcal{H}_N}$ as
\begin{align*}
    P_{\Psi,\Phi} \mapsto \mathrm{dens}(P_{\Psi,\Phi}) = \rho_{\Psi,\Phi} \in \mL^1(I).
\end{align*}
Since (by Cauchy-Schwarz)
\begin{align*}
    \norm{\rho_{\Psi,\Phi}}_{\mL^1} \leq N  \norm{\Psi}_{\mathcal{H}_N} \norm{\Phi}_{\mathcal{H}_N} = N \mathrm{Tr} |P_{\Psi,\Phi}|,
\end{align*}
the density map can be continuously extended by linearity to the space of trace class operators. Moreover, this extension satisfies the duality relation
\begin{align}
    \mathrm{Tr} \, A^\ast \hat{v} = \int_I \rho_{A^\ast}(x) v(x)  \mathrm{d} x = \int_I \overline{\rho_{A}(x)} v(x) \mathrm{d} x \quad \mbox{for any $v\in \mL^\infty(I;\C)$ and $A \in \mathcal{S}_1(\mathcal{H}_N)$.} \label{eq:density duality L1}
\end{align}

In a similar way, for any $\Gamma \in \mathcal{S}_1^1$, we can define the associated complex-valued density $\rho_\Gamma \in \mH^1(I)$ via the Riesz representation theorem. More precisely, by the Riesz representation theorem in $\mH^{-1}(I)$, there exists a unique element $\rho_\Gamma \in \mH^1(I)$ such that
\begin{align}
    \inner{\rho_\Gamma,v} = \mathrm{Tr} \{ \Gamma^\ast \hat{v} \} = \mathrm{Tr}  \left\{(\mathcal{C}_\Delta \Gamma)^\ast (\mathcal{C}_\Delta^{-1}\hat{v})\right\} ,\quad \mbox{for any $v \in \mH^{-1}(I)$.} \label{eq:duality density definition}
\end{align}
In particular, this yields a $\C$-linear continuous map $\mathrm{dens} : \mathcal{S}_1^1 \rightarrow \mH^1(I)$ sending an operator $\Gamma$ to its density. Note that, by the duality relation in~\eqref{eq:density duality L1}, this map agrees with the density map defined for trace class operators. Moreover, it is not difficult to check that for positive $\Gamma$, the density $\rho_\Gamma$ is just the usual single-particle density of the density matrix $\Gamma$.

Let us now introduce the following operator:
\begin{align}
    B_\Psi : \mathcal{H}_N^1  \rightarrow \mH^1(I),\quad  \Phi \mapsto B_\Psi(\Phi) = N \int_{I_{N-1}} \overline{\Psi(x,x_2,...,x_N)} \Phi(x,x_2,...,x_N) \mathrm{d} x_2... \mathrm{d} x_N. \label{eq:Bdef}
\end{align}
The real part of this operator can be seen as the Fr\'echet differential at $\Psi$ of the wave-function to density map, $\Phi \mapsto \mathrm{dens}(\Phi)$, i.e., 
\begin{align*}
    \mathrm{dens}(\Psi+\Phi) = \mathrm{dens}(\Psi) + 2 \mathrm{Re} \, B_\Psi \Phi +\mathrm{dens}(\Phi),
  \qquad
  \norm{\mathrm{dens}(\Phi)}_{\mH^1(I)}
  \lesssim \norm{\Phi}_{\mathcal{H}_N^1}^2. 
\end{align*}
Moreover, let us denote by $B_\Psi^\ast : \mH^{-1}(I;\C) \rightarrow \mathcal{H}_N^{-1}$ the adjoint operator, which acts as
\begin{align}
    \inner{\Phi,B_\Psi^\ast v} = \inner{B_\Psi \Phi, v} = \inner{\rho_{\Phi,\Psi}, v} = \inner{\Phi,\hat{v} \Psi}.\label{eq:B adjoint def}
\end{align}
The adjoint operator can also be restricted to $\mH^1(I)$. In this case, it defines a bounded operator from $\mH^1(I)$ to $\mathcal{H}^1_N$, as shown next. 
\begin{proposition}[Adjoint operator] Let $\Psi \in \mathcal{H}_N^1$. Then for any $f \in \mH^1(I)$, we have
\begin{align}
    (B_\Psi^\ast f)(x_1,...,x_N) = \sum_{j=1}^N f(x_j) \Psi(x_1,...,x_N) .  \label{eq:Badjoint}
\end{align}
Moreover, $B_\Psi^\ast : \mH^1(I) \rightarrow \mathcal{H}^1_N$ is bounded. In particular, the operator $B_{\Psi}$ extends to a bounded linear operator $B_{\Psi}:\mathcal{H}_N^{-1} \rightarrow \mH^{-1}(I)$.
\end{proposition}
\begin{proof} Since $\mH^1(I) \subset \mL^\infty(I)$, the function $\sum_{j} f(x_j) \Psi$ is clearly square integrable. To see that it has finite kinetic energy, note that
\begin{align*}
    \partial_{x_i} (B_\Psi^\ast f) = \partial_{x_i} f(x_i) \Psi + \sum_{j} f(x_j) \partial_{x_i} \Psi .
\end{align*}
Again, the second term is square integrable because $f\in \mL^\infty(I)$. For the first term we have
\begin{align*}
    \int_{I_N} |\partial_x f(x)|^2 |\Psi(x,x_2,...,x_N)|^2\mathrm{d} x \mathrm{d} x_2 ... \mathrm{d} x_N = \frac{1}{N} \int_{I} |\partial_x f(x)|^2 \rho_\Psi(x) \mathrm{d} x.
\end{align*}
Since $\rho_{\Psi} \in \mH^1(I) \subset \mL^\infty(I)$, this term is also square integrable.
\end{proof}
The operators $B_\Psi$ and $B_{\Psi}^\ast$ will play a fundamental role in the proofs of the next section. 

 Finally, we recall that $\mathcal{X}_0$ is defined as the space of real-valued $\mH^1$ functions with average $0$:
\begin{align*}
    \mathcal{X}_0 = \left\{ f\in \mH^1(I;\R) : \int_I f(x) \mathrm{d} x = 0\right\}.
\end{align*}
As mentioned in Remark~\ref{rem:tanget space}, this corresponds to the space of density variations.

\section{A quantitative version of the Hohenberg--Kohn theorem}
\label{sec:QHK proof}
In this section, our goal is to prove Theorem~\ref{thm:QHK}. For this proof, we shall use a few auxiliary lemmas. The first lemma is a standard estimate showing that, if a wavefunction has energy close to the ground-state energy, then it must be $\mH^1$-close to the ground-state wavefunction.
\begin{lemma}[Variational principle] \label{lem:VP} Let $v\in \mathcal{V}(I)$, $w\in \mathcal{W}(I)$, $N\in \N$ and $\Psi$ be the normalized ground-state of the operator $H_N(v,w)$. Let $E_0 = \min \sigma(H_N(v,w))$ be the ground-state energy. Then, there exists a constant $C = C(v,w)$ such that, for any normalized state $\Psi' \in \mathcal{H}_N^1$, there exists a global phase $\alpha \in \C$, $|\alpha| =1$ such that
\begin{align}
    \norm{\Psi'-\alpha \Psi}_{\mH^1(I_N)} \leq C \sqrt{\mathfrak{a}_{v,w}(\Psi')- E_0}. \label{eq:wavefunction bound}
\end{align}
\end{lemma}

\begin{proof} To simplify the notation, we denote the excess energy of $\Psi'$ by 
\begin{align}
    \epsilon \coloneqq \mathfrak{a}_{v,w}(\Psi') - E_0 \geq 0. \label{eq:excess energy}
\end{align}
Then note that
    \begin{align*}
        E_0 |\inner{\Psi, \Psi'}|^2 + E_1 \norm{P_\Psi^\perp \Psi'}^2 \leq \mathfrak{a}_{v,w}(\Psi') = E_0 + \epsilon = E_0(|\inner{\Psi,\Psi'}|^2 + \norm{P_\Psi^\perp \Psi'}^2) + \epsilon
    \end{align*}
    where $E_1$ is the energy of the first excited state of $H_N(v,w)$. Thus, 
    \begin{align}
        \norm{P_\Psi^\perp \Psi'}^2 \leq \epsilon/\omega_1, \label{eq:simple est}
    \end{align}
    where $\omega_1 = E_1 - E_0>0$ is the first excitation energy of $H_N(v,w)$. Since $\Psi'$ is normalized we can assume without loss of generality that $\epsilon>0$ is small. Indeed, note that
    \begin{align}
        \norm{\Psi' - \alpha \Psi}_{\mH^1} \leq \norm{\Psi'}_{\mH^1} + \norm{\Psi}_{\mH^1} \lesssim \sqrt{\mathfrak{a}_{v,w}(\Psi') + (1-E_0) \norm{\Psi'}_{\mL^2}} + \norm{\Psi}_{\mH^1} = \sqrt{\epsilon +1} + \norm{\Psi}_{\mH^1}, \label{eq:previous est}
    \end{align}
    where we used the equivalence between the $\mathcal{H}_N^1$ norm and the form norm (see~\eqref{eq:norm equivalence}) in the second inequality. From~\eqref{eq:previous est}, we see that~\eqref{eq:wavefunction bound} holds for $\epsilon \geq \epsilon_0$ with $C$ given by the implicit constant in estimate~\eqref{eq:previous est} times
    \begin{align*}
        C(\epsilon_0) \coloneqq \sup_{\epsilon > \epsilon_0} \frac{\sqrt{\epsilon+1} + \norm{\Psi}_{\mH^1}}{\sqrt{\epsilon}} < \infty.
    \end{align*}
        Hence, we can restrict our attention to $\epsilon < \epsilon_0$ for some $\epsilon_0>0$ fixed. So let us set $\epsilon_0 = \omega_1$, where $\omega_1$ is the first excitation energy of $H_N(v,w)$. Then, we find from~\eqref{eq:simple est} that
    \begin{align}
        1\geq |\inner{\Psi,\Psi'}| = \sqrt{1-\norm{P_{\Psi}^\perp  \Psi'}^2} \geq \sqrt{1- \epsilon/\omega_1} > 1-\epsilon/\omega_1. \label{eq:firstests}
    \end{align}
    In particular, the inner product $\inner{\Psi,\Psi'}$ is nonzero and we can define
    \begin{align*}
        \alpha \coloneqq \frac{\inner{\Psi,\Psi'}}{|\inner{\Psi,\Psi'}|}.
    \end{align*}
    Then, from~\eqref{eq:firstests},
    \begin{align}
        \norm{\Psi'-\alpha \Psi}^2 = 2 - 2 \mathrm{Re}\{ \alpha \inner{\Psi',\Psi}\} = 2- 2|\inner{\Psi',\Psi}|\leq 2\epsilon/\omega_1 . \label{eq:L2est}
    \end{align}
    To conclude, we can use the equivalence between the $\mH^1$ and the $\mathfrak{a}_{v,w}$ norms together with the definition of excess energy in~\eqref{eq:excess energy}, the ground-state identity $\mathfrak{a}_{v,w}(\Psi,\Psi') = E_0 \inner{\Psi,\Psi'}$, and~\eqref{eq:L2est} to obtain
    \begin{align*}
        \norm{\Psi'-\alpha \Psi}_{\mH^1}^2 \lesssim_{v,w} \norm{\Psi'-\alpha \Psi}_{\mathfrak{a}_{v,w}}^2 &= \mathfrak{a}_{v,w}(\Psi) + \mathfrak{a}_{v,w}(\Psi') - 2E_0\mathrm{Re}\{ \alpha \inner{\Psi',\Psi}\} + (1-E_0)\norm{\Psi'-\alpha \Psi}^2 \\
        &\leq E_0 (2-2|\inner{\Psi,\Psi'}|) + (1-E_0) \norm{\Psi'-\alpha \Psi}^2 +\epsilon \\
        &= \norm{\Psi'-\alpha \Psi}^2 +\epsilon \leq \left(1+\frac{2}{\omega_1}\right) \epsilon, 
    \end{align*}    which concludes the proof. 
\end{proof}

\begin{remark}[Degenerate case] Note that, in the case of a degenerate ground-state, the result from Lemma~\ref{lem:VP} still holds, but with $\alpha \Psi$ replaced by the normalized projection of $\Psi'$ on the ground-state space of $H_N(v,w)$. 
\end{remark}

The next lemma shows that the representing potential function $\rho \mapsto v(\rho;w)$ is locally bounded.

\begin{lemma}[Locally uniform bound on representing potential]\label{lem:unif bound potential}
Let $\rho \in \mathcal{D}_N$, then there exists a neighborhood $\rho \in U\subset \mathcal{D}_N$ and a constant $C = C(\rho)>0$ such that 
\begin{align}
    \norm{v(\rho';w)}_{\mathcal{V}/\{1\}} \leq C, \quad \mbox{for any $\rho' \in U$.} \label{eq:uniform potential control}
\end{align}
Moreover, if $v' \in v(\rho';w)$ is the representative such that $H_N(v',w)$ has zero ground-state energy, then
\begin{align}
    \norm{v'}_{\mH^{-1}} \mathrel{\lesssim_\rho} 1, \quad \mbox{for any $\rho'\in U$,} \label{eq:really uniform potential control}
\end{align}
with an implicit constant that may depend on $\rho$ but is independent of $\rho'$. 
\end{lemma}

\begin{proof}
The proof here uses two important facts. First, the potential $v(\rho;w)\in \mathcal{V}/\{1\}$ is, up to a sign, the subgradient of the Levy--Lieb functional $F_{\rm LL}(\rho;w)$ defined in~\eqref{eq:LL def}. Second, this functional satisfies the bounds
\begin{align}
    -1 \lesssim F_{\rm LL}(\rho;w) \lesssim \norm{\sqrt{\rho}}_{\mH^1}^2, \label{eq:lower + upper bound}
\end{align}
with implicit constants that may depend on $w$ and $N$ but are independent of $\rho$. 

The first fact was proved in \cite[Lemma 6.1]{Cor26a}, and we refer to that paper for the proof. The upper bound in~\eqref{eq:lower + upper bound} follows from a well-known $N$-representability upper bound due to Lieb \cite[Theorem 1.2]{Lie83} (see also \cite[Lemma 4.6]{Cor26a} for the simpler estimate in the current 1D setting) and the fact that the interaction is Laplace bounded in the sense of forms. The lower bound holds because the interaction $w$ is infinitesimally Laplace bounded and the kinetic energy is non-negative. Indeed, using the fact that $w$ is Laplace bounded with bound $<1/2$, we get 
\begin{align*}
    \inner{\Psi, H_N(0,w) \Psi} = \norm{\nabla \Psi}_{\mL^2}^2 + \inner{\rho^{(2)}_\Psi,w} \geq \frac12 \norm{\nabla \Psi}_{\mL^2}^2 - C \geq - C
\end{align*}
for some constant $C>0$ depending on the norm $\norm{w}_{\mW^{-1,p}}$. (Note that, if the interaction is positive, i.e., repulsive, then it can be completely neglected for the lower bound and $F_{\rm LL}(\rho;w) \geq 0$.)

We can now combine these facts to prove the lemma as follows. Let $\rho \in \mathcal{D}_N$ and $\epsilon= \epsilon(\rho)>0$ be so small that $\rho + \delta \in \mathcal{D}_N$ and 
\begin{align}
    \norm{\sqrt{\rho+\delta}}_{\mH^1} \leq C \quad \mbox
\ \quad \mbox{for any $\delta \in B_{2\epsilon}(0) \subset \mathcal{X}_0$ and some $C>0$.} \label{eq:uniform upper bound}
\end{align}
Note that explicit bounds on $\epsilon$ and $C>0$ can be found, for instance, in terms of the smallest value $c = \inf \rho>0$ and the implicit constant in the Sobolev inequality $\norm{f}_{\mL^\infty} \lesssim \norm{f}_{\mH^1}$. Then, since $v(\rho';w)$ is the subgradient of $F_{\rm LL}(\rho';w)$, we have 
\begin{align*}
    F_{\rm LL}(\rho';w) - \inner{\delta, v(\rho';w)} \leq F_{\rm LL}(\rho'+\delta).
\end{align*}
As $v$ is a real-valued distribution, we can now use ~\eqref{eq:lower + upper bound} and~\eqref{eq:uniform upper bound} to obtain 
\begin{align*}
    \epsilon \norm{v(\rho';w)}_{\mathcal{V}/\{1\}} = \sup_{\delta \in B_\epsilon(0)} -\inner{\delta,v(\rho';w)} \lesssim (1+\norm{\sqrt{\rho'}}_{\mH^1}) \lesssim 1+C,
\end{align*}
for any $\norm{\rho'-\rho}_{\mH^1} \leq \epsilon$. This proves~\eqref{eq:uniform potential control} with $U = B_\epsilon(\rho)$.

To prove the last estimate, let $\tilde{v}\in v(\rho';w)$ be a representative  such that 
\begin{align}
    \norm{\tilde{v}}_{\mH^{-1}} \leq 2 \norm{v(\rho',w)}_{\mH^{-1}/\{1\}}. \label{eq:some est}
\end{align}
Since the ground-state energy of $H_N(\tilde{v},w)$ is given by
\begin{align*}
    E(\tilde{v};w)  = F_{\rm LL}(\rho';w) + \inner{\rho',\tilde{v}},
\end{align*}
the representative $v' \in v(\rho';w)$ with ground-state energy zero is given by
\begin{align*}
    v' = \tilde{v} - E(\tilde{v};w)/N = \tilde{v} - \frac{F_{\rm LL}(\rho';w) + \inner{\rho',\tilde{v}}}{N}.
\end{align*}
We can thus use estimates~\eqref{eq:uniform potential control},~\eqref{eq:lower + upper bound}, and~\eqref{eq:some est} to obtain the uniform bound~\eqref{eq:really uniform potential control} in $U= B_\epsilon(\rho)$. 
\end{proof}

For the proof of Theorem~\ref{thm:QHK}, we need one additional lemma, which is the key lemma in the paper.

\begin{lemma}[Invertibility of $B_\Psi B_\Psi^\ast$] \label{lem:inverse} Let $\Psi$ be the ground-state of $H_N(v,w)$. Then the map $B_\Psi B^\ast_\Psi\frac{1}{\rho_\Psi} : \mathcal{X}_0\rightarrow \mathcal{X}_0$ is invertible in $\mathcal{B}(\mathcal{X}_0)$, where $1/\rho_\Psi$ denotes the operator of multiplication by $1/\rho_\Psi$. In particular, the operator $B_\Psi : \{\Psi\}^\perp \cap \mathcal{H}_N^1 \rightarrow \mathcal{X}_0$ is surjective. 
\end{lemma}

\begin{remark}[Lagrange multiplier] The surjectivity of $B_\Psi$ shows that every admissible infinitesimal density variation can be realized by an admissible variation of the wave‑function.
Note that this is a sufficient regularity condition for the existence of a Lagrange multiplier for the constrained optimization in $F_{\mathrm{LL}}(\rho)$, see \cite[Chapter 9]{Lue69}. A similar strategy was employed to investigate the constrained-search functional (and existence of Lagrange multipliers) in a simplified model in the QEDFT setting \cite{BCL+25}. 
\end{remark}

To prove this lemma, we shall need some properties of the operator $K$ defined as
\begin{align}
    K(x,y) = \frac{\rho_\Psi^{(2)}(x,y)}{\rho_\Psi(y)}, \label{eq:Kdef}
\end{align}
where $\Psi$ is the ground-state of $H_N(v,w)$.
\begin{lemma}[The pair density operator] \label{lem:pair density operator} The operator $K$ defined in~\eqref{eq:Kdef} is a bounded operator in $\mathcal{B}(\mL^2_0(I))$ (where $\mL^2_0$ is the space of real-valued $\mL^2$ functions with zero mean) and in $\mathcal{B}(\mathcal{X}_0)$. Moreover $K$ is also bounded from $\mL^\infty(I)$ to $\mH^1(I)$ and from $\mL^2_0(I)$ to $\mW^{1,1}(I)$, i.e., it satisfies
\begin{align*}
    \norm{Kf}_{\mH^1} \lesssim \norm{f}_{\mL^\infty} \quad\mbox{and}\quad \norm{Kf}_{\mW^{1,1}} \lesssim \norm{f}_{\mL^2},
\end{align*}
with implicit constants independent of $f$.
\end{lemma}

\begin{proof}
    Since $\int_I \rho_\Psi^{(2)}(x,y) \mathrm{d} x = (N-1) \rho_\Psi(y)$, it is easy to see that $K$ maps functions with zero mean to functions with zero mean. Moreover, since $\mH^1 \subset \mW^{1,1} \subset \mL^\infty \subset \mL^2$ by the Sobolev embedding, to show that $K\in \mathcal{B}(\mL^2_0(I)) \cap \mathcal{B}(\mathcal{X}_0)$ it suffices to prove that $K$ maps bounded functions to $\mH^1$ functions and $\mL^2$ functions to $\mW^{1,1}(I)$ functions. The $\mL^\infty(I)$ to $\mH^1(I)$ bound is proven in \cite[Lemma 5.1]{Cor26a} and we refer to that paper for the proof. For the $\mW^{1,1}$ bound, we note that by Fubini and two times Cauchy-Schwarz we have
    \begin{align*}
        \norm{\partial_x (Kf)}_{\mL^1(I)} &= 2 (N-1) N\int_I \left| \int_I \left(\int_{I_{N-2}} \mathrm{Re}\{\partial_x \Psi(x,y,x_3,...) \overline{\Psi(x,y,x_3,...)}\} \mathrm{d}x_3... \mathrm{d} x_N\right) \frac{f(y)}{\rho(y)} \mathrm{d} y \right|\mathrm{d} x \\
        & \leq 2(N-1)\sqrt{N} \int_I \frac{|f(y)|}{\rho(y)} \rho(y)^{\frac12} \left(\int_{I_{N-1}} |\partial_x \Psi(x,y,...,x_N)|^2 \mathrm{d} x\mathrm{d} x_2... \mathrm{d} x_N\right)^{\frac12} \mathrm{d} y\\
        &\leq 2(N-1) \sqrt{N} \norm{1/\rho}_{\mL^\infty(I)}^{\frac12} \norm{f}_{\mL^2(I)} \norm{\partial_x \Psi}_{\mL^2(I_N)},
    \end{align*}
    which proves the desired bound. The bound $\norm{Kf}_{\mL^1} \lesssim \norm{f}_{\mL^2}$ can be proved in a similar way (or using the derivative bound and the Poincar\'e inequality since the range of $K$ has zero mean).
\end{proof}

\begin{proof}[Proof of Lemma~\ref{lem:inverse}] First, from~\eqref{eq:Bdef},~\eqref{eq:Badjoint}, and straightforward calculation we find
\begin{align}
    \left(B_\Psi B_\Psi^\ast \frac{1}{\rho_\Psi} f\right)(x) = f(x) + \int_I \rho^{(2)}_\Psi(x,y) \frac{f(y)}{\rho_\Psi(y)} \mathrm{d} y = \left((I+K) f\right)(x),\label{eq:id0}
\end{align}
where $K$ is the operator with integral kernel defined in~\eqref{eq:Kdef}.

Next, we show that $I+K$ is injective in $\mL^2_0$. To this end, we first note that
\begin{align*}
    \inner{f, B_\Psi B_\Psi^\ast f}_{\mL^2(I)} = \norm{B_\Psi^\ast f}_{\mL^2(I_N)}^2.
\end{align*}
As the multiplication operator $1/\rho_{\Psi}$ is an isomorphism in both $\mL^2(I)$ and $\mH^1(I)$ (recall that $\rho_\Psi \geq c$ for some $c>0$), the above equation implies that $f \in \ker (I+K)$ if and only if 
\begin{align*}
    \left(B_\Psi^\ast \frac{1}{\rho_\Psi}f\right)(x) = \sum_{j=1}^N \frac{f(x_j)}{\rho_\Psi(x_j)} \Psi(x_1,...,x_N) = 0 \quad \mbox{for a.e. $(x_1,...,x_N) \in I_N$.} 
\end{align*}
However, from the unique continuation proved in \cite[Theorem 2.1]{Cor26b}, we have $\Psi \neq 0$ a.e., which implies that $\sum_{j=1}^N f(x_j)/\rho_\Psi(x_j) = 0$. This in turn implies that $f = 0$. Hence, we have shown that $I+K$ is injective in $\mL^2_0(I)$ and therefore also injective in $\mathcal{X}_0$. To conclude the proof, by the bounded inverse theorem, it suffices to show that $I+K$ is surjective in $\mathcal{B}(\mathcal{X}_0)$.

For this, first note that $K$ is compact as an operator in $\mathcal{B}(\mL^2_0)$. Indeed, this follows from the fact that $K$ maps $\mL^2_0$ boundedly to $\mW^{1,1}\cap \mL^2_0$ and the latter is compactly embedded (by the Sobolev embedding) in $\mL^2$. Since $K$ is compact and $I+K$ is injective, we can apply the Fredholm alternative (see, e.g., \cite[Theorem VII.7.9 and Corollary VII.7.10]{Con90}) to conclude that $B_\Psi B_\Psi^\ast1/\rho_\Psi = I +K$ is invertible in $\mL^2_0$. We can now prove that $I+K$ is surjective in $\mathcal{B}(\mathcal{X}_0)$ by using the regularity improving property of $K$. Precisely, let $g\in \mathcal{X}_0 \subset \mL^2_0(I)$ and set $f \coloneqq (I+K)^{-1} g \in \mL^2_0(I)$. Since $K$ maps $\mL^2$ to $\mW^{1,1} \subset \mL^\infty$ and $\mL^\infty$ to $\mH^1$, we have
\begin{align*}
    K^2 f \in \mathcal{X}_0\quad \mbox{for any $f\in \mL^2_0(I)$. }
\end{align*}
Hence, $Kf = Kg - K^2 f \in \mathcal{X}_0$, and therefore $f = g - Kf \in \mathcal{X}_0$. Thus, $(I+K)^{-1} g\in \mathcal{X}_0$ for any $g\in \mathcal{X}_0$. This shows that $I+K$ is surjective in $\mathcal{X}_0$, which concludes the proof.
\end{proof}

We can now proceed with the proof of Theorem~\ref{thm:QHK}.

\begin{proof}[Proof of Theorem~\ref{thm:QHK}]
Let $\rho,\rho' \in \mathcal{D}_N$, then we denote by $v$, respectively $v'$, a representative of the class $v(\rho;w)$, respectively $v(\rho';w)$. Moreover, let $\Psi$ and $\Psi'$ be the ground-states of $H_N(v,w)$ and $H_N(v',w)$.

So first, we use the variational principle to obtain
\begin{align*}
    \mathfrak{a}_{v,w}(\Psi') &= \mathfrak{a}_{v',w}(\Psi') + \inner{\rho',v-v'}\leq \mathfrak{a}_{v',w}(\Psi) +\inner{\rho',v-v'} = \mathfrak{a}_{v,w}(\Psi) + \inner{\rho'-\rho,v-v'}. 
\end{align*}
Since $\rho-\rho'$ has zero mean (i.e., belongs to $\mathcal{X}_0$), the last term is invariant under adding or subtracting a constant from $v$ or $v'$. Therefore, 
\begin{align*}
    \mathfrak{a}_{v,w}(\Psi') \leq \mathfrak{a}_{v,w}(\Psi) + \norm{v-v'}_{\mathcal{V}(I)/\{1\}} \norm{\rho'-\rho}_{\mH^1}.
\end{align*}
As $\Psi$ is the ground-state of $H_N(v,w)$, by Lemma~\ref{lem:VP} we find that (up to a constant phase factor),
\begin{align}
    \norm{\Psi - \Psi'}_{\mH^1}^2 \lesssim_{v,w} \norm{v-v'}_{\mathcal{V}(I)/\{1\}} \norm{\rho-\rho'}_{\mH^1}. \label{eq:H1est}
\end{align}
Moreover, after possibly shifting the potentials $v$ and $v'$ by a constant, we can assume that both ground-state energies are zero. Consequently, we have
\begin{align}
    0 = \mathfrak{a}_{v',w}(\Phi,\Psi')- \mathfrak{a}_{v,w}(\Phi,\Psi) &= \mathfrak{a}_{v',w}(\Phi,\Psi'-\Psi) + \mathfrak{a}_{v',w}(\Phi,\Psi) - \mathfrak{a}_{v,w}(\Phi,\Psi) \nonumber \\
    &= \mathfrak{a}_{v',w}(\Phi,\Psi'-\Psi) + \inner{B_\Psi \Phi, v'-v}. \label{eq:sime middle eq}
\end{align}
We now note that, for any $\rho'$ in a sufficiently small neighborhood of $\rho$, by Lemma~\ref{lem:unif bound potential} we have the uniform bound $\norm{v'}_{\mH^{-1}} = \norm{v(\rho';w)}_{\mH^{-1}} \leq C$ with a constant that may depend on $\rho$ but is independent of $\rho'$. Consequently, the first term in~\eqref{eq:sime middle eq} can be estimated by~\eqref{eq:uniform potential control} and~\eqref{eq:H1est} as
\begin{align*}
    |\mathfrak{a}_{v',w}(\Psi-\Psi',\Phi)| \lesssim \norm{\Psi-\Psi'}_{\mH^1} \norm{\Phi}_{\mH^1} \lesssim \norm{v-v'}_{\mathcal{V}(I)/\{1\}}^{\frac12} \norm{\rho-\rho'}_{\mH^1}^{\frac12} \norm{\Phi}_{\mH^1},
\end{align*}
with a constant independent of $v'$ (provided that $\rho'$ lies in a sufficiently small but fixed neighborhood of $\rho$). Using this bound back in~\eqref{eq:sime middle eq}, we conclude that
\begin{align}
    |\inner{B_{\Psi} \Phi, v'-v}| =|\mathfrak{a}_{v',w}(\Phi,\Psi'-\Psi)| \lesssim \norm{\Phi}_{\mH^1} \norm{v'-v}_{\mathcal{V}(I)/\{1\}}^{\frac12} \norm{\rho'-\rho}_{\mH^1}^{\frac12},  \label{eq:niceest}
\end{align}
for any $\Phi\in \mathcal{H}_N^1$. 

The result now follows from the following observation: since $B_{\Psi} B_{\Psi}^\ast\frac{1}{\rho_\Psi}$ is invertible in $\mathcal{B}(\mathcal{X}_0)$ by Lemma~\ref{lem:inverse}, we can test against enough functions $\Phi$ to estimate the norm of $v'-v$. More precisely, by Lemma~\ref{lem:inverse}, the image of all $\Phi = B_{\Psi}^\ast \frac{1}{\rho_{\Psi}}f$ with $f\in \mathcal{X}_0$ and $\norm{f}_{\mH^1} \leq 1$ under $B_\Psi$ contains a ball centered at the origin in $\mathcal{X}_0$ with some radius $\delta>0$. Therefore, by~\eqref{eq:niceest}
\begin{align*}
    \norm{v'-v}_{\mathcal{V}(I)/\{1\}} = \frac{1}{\delta} \sup_{\substack{g\in \mathcal{X}_0\\ \norm{g}\leq \delta}} |\inner{g,v'-v}|\leq \frac{1}{\delta} \sup_{\substack{f\in \mathcal{X}_0\\ \norm{f}\leq 1}} \left|\left\langle B_{\Psi}B_{\Psi}^\ast\frac{f}{\rho_{\Psi}}, v'-v\right\rangle \right|\lesssim \norm{v'-v}_{\mathcal{V}(I)/\{1\}}^{\frac12}\norm{\rho'-\rho}_{\mH^1}^{\frac12},
\end{align*}
which completes the proof.
\end{proof}

\section{Analyticity of the density-to-potential map} \label{sec:analytic proof}

In this section, our goal is to prove Theorem~\ref{thm:analytic map}. 
As previously mentioned, the key idea in the proof is to study the potential-to-density map $\rho :\mathcal{V}(I)/\{1\}\times \R \rightarrow \mathcal{D}_N$ sending a potential to the density $\rho(v,\lambda)$ of the unique ground-state of $H_N(v,\lambda w)$. 

To this end, we shall study the following nonlinear operators. For fixed $N\in \N$ and $w\in \mathcal{W}(I)$, we define 
\begin{align}
    P: \mathcal{V}/\{1\} \times \R\rightarrow \mathcal{S}_1^1 \quad \mbox{as the mapping}\quad (v,\lambda) \mapsto P(v,\lambda), \label{eq:projection-valued map}
\end{align}
where $P(v,\lambda)$ is the $\mL^2$-orthogonal projection on the ground-state space of $H_N(v, \lambda w)$ and $\mathcal{S}_1^1$ is the trace-class space introduced in Section~\ref{sec:trace spaces}. Moreover, we introduce the mapping
\begin{align*}
    R^\perp : \mathcal{V}/\{1\}\times \R \rightarrow \mathcal{B}_{-1,1}\quad (v,\lambda) \mapsto R^\perp(v,\lambda),
\end{align*}
where $R^\perp(v,\lambda)$ is the reduced resolvent of $H_N(v,\lambda w)$ introduced in~\eqref{eq:reduced resolvent}. Note that both $P(v,\lambda)$ and $R^\perp(v,\lambda)$ are invariant under a constant shift of the potential; in particular, these maps are well-defined in the quotient space $\mathcal{V}/\{1\}$.

The first step here is to show that the map $P$ is analytic on the appropriate function spaces and obtain an explicit expression for its derivative.
\begin{lemma}[Analytic projection map] \label{lem:smooth projection} The map $(v,\lambda) \mapsto P(v,\lambda)$ is in $C^\omega(\mathcal{V}/\{1\}\times \R; \mathcal{S}_1^1)$. Moreover, the (Fr\'echet) derivative of $P$ at $(v,\lambda)$ is given by
\begin{align}
    D_{v,\lambda} P(u,\mu) = R^\perp(v,\lambda)(\hat{u}+\mu \widehat{w})\, \, P(v,\lambda) + P(v,\lambda) (\hat{u}+\mu \widehat{w})\, \,R^\perp(v,\lambda), \quad (u,\mu) \in \mathcal{V}/\{1\}\times \R, \label{eq:derivative formula}
\end{align}
where $\hat{u} =\sum_{j} u_j$ and $\hat{w} = \sum_{j\neq k} w_{j,k}$ are the $N$-particle operators defined in~\eqref{eq:quantized operators}.
\end{lemma}

\begin{proof} The proof here is essentially an application of resolvent formulas. To simplify the notation, we shall occasionally omit the arguments $(v,\lambda)$ and simply write $P$ and $R^\perp$ for $P(v,\lambda)$ and $R^\perp(v,\lambda)$. 

We start by noticing that, since $P(v+\alpha,\lambda) = P(v,\lambda)$ for any $\alpha \in \R$, it suffices to show that the map $(v,\lambda) \in \mathcal{V}\times \R \mapsto P(v,\lambda)$ is analytic. For this, we first recall (see \cite[Theorem 2.1]{Cor26b}) that the ground-state energy of $H_N(v,\lambda w)$ is a simple isolated eigenvalue. Thus, assuming without loss of generality that the ground-state energy is zero, we can find $\delta>0$ so small that $\mathcal{C} = \partial B_\delta(0)$ encloses no excitation energy of $H_N(v,\lambda w)$. Moreover, from standard perturbation theory, the same contour encloses only the ground-state energy of $H_N(v+u, (\lambda + \mu)w)$ provided that $\norm{u}_{\mathcal{V}}$ and $|\mu|$ are small enough. Therefore, using the well-known resolvent (or Neumann) series
\begin{align*}
    R_z(v+u,\lambda+\mu) \coloneqq (z-H_N(v+u,(\lambda+\mu)w))^{-1} = R_z(v,\lambda) \sum_{n\geq 0} \left((\hat{u}+\mu \hat{w}) R_z(v,\lambda)\right)^n,
\end{align*}
we find that
\begin{align*}
    P(v+u,\lambda+\mu) = \frac{1}{2 \pi \ii} \oint_{\partial B_\delta(0)} R_z(v+u,\lambda+\mu) \mathrm{d} z = \sum_{n\geq 0} \frac{1}{2 \pi \ii} \oint_{\partial B_\delta(0)} R_z(v,\lambda) \left((\hat{u}+\mu \widehat{w}) R_z(v,\lambda)\right)^n \mathrm{d} z.
\end{align*}
We can now use the spectral decomposition
\begin{align}
    R_z(v,\lambda) = \frac{P}{z} +R^\perp_z(v,\lambda), \quad\mbox{where} \quad R^\perp_z(v,\lambda) = \sum_{j\geq 1} \frac{P_j(v,\lambda)}{z-\omega_j}, \label{eq:spectral decomposition}
\end{align}
and the residue theorem to find that
\begin{align*}
    P(v+u,\lambda+\mu) = P + R^\perp (\hat{u} +\mu \widehat{w}) P + P (\hat{u} +\mu \widehat{w}) R^\perp + \sum_{n\geq 2} \mathbb{P}_n\left((u,\mu),...,(u,\mu)\right),
\end{align*}
where $\mathbb{P}_n : (\mathcal{V}\times \R) ^n \rightarrow \mathcal{B}_{-1,1}$ are the $n$-linear operators
\begin{align*}
    \mathbb{P}_n\left((u,\mu),...,(u,\mu)\right)   =\frac{1}{ 2\pi \ii} \oint_{\partial B_\delta(0)} R_z(v,\lambda) \left((\hat{u}+\mu \widehat{w}) R_z(v,\lambda)\right)^n \mathrm{d} z.
\end{align*}
Therefore, according to Definition~\ref{def:analytic}, the map $P$ is analytic from $\mathcal{V}\times \R$ to $\mathcal{S}_1^1$ if and only if we can show an estimate of the form
\begin{align}
    \norm{\mathbb{P}_n\left((u,\mu)^n\right)}_{\mathcal{S}_1^1} \leq R^n (\norm{u}_{\mathcal{V}}+|\mu|)^n \quad \mbox{for some $R>0$ and any $n\geq 1$,} \label{eq:analytic goal}
\end{align}
where we abbreviated $\mathbb{P}_n\left((u,\mu),...,(u,\mu)\right)$ to $\mathbb{P}_n\left((u,\mu)^n\right)$. To establish this estimate, we claim it suffices to show that the rank of $\mathbb{P}_n\left((u,\mu)^n\right)$ grows at most exponentially with $n$. Indeed, if we can show that $\mathrm{rank} \left(\mathbb{P}_n\left((u,\mu)^n\right)\right) \leq C^n$ with a constant independent of $u$ and $\mu$, then from the simple estimates
\begin{align*}
    \norm{T}_{\mathcal{S}_1^1} \leq \mathrm{rank}(T) \norm{T}_{-1,1}, \quad \norm{AB}_{-1,1} \leq \norm{A}_{-1,1} \norm{B}_{-1,1},\quad \mbox{and}\quad M \coloneqq \max_{z\in \partial B_\delta(0)} \norm{R_z(v,\lambda)}_{-1,1} \leq \infty,
\end{align*}
we find that
\begin{align*}
    \norm{\mathbb{P}_n\left((u,\mu)^n\right)}_{\mathcal{S}_1^1} &\leq C^n \norm{\mathbb{P}_n\left((u,\mu)^n\right)}_{-1,1} \leq \frac{C^n}{2\pi} \oint_{\partial B_\delta(0)}  \norm{R_z(v,\lambda) \left((\hat{u} +\mu \widehat{w})R_z(v,\lambda)\right)^n}_{-1,1} \mathrm{d} |z| \\
    &\leq C^n M^{n+1} \delta (N\norm{u}_{\mathcal{V}}+\mu\norm{\widehat{w}}_{-1,1})^n, 
\end{align*}
which implies~\eqref{eq:analytic goal}. 

To obtain a bound on the rank of $\mathbb{P}_n\left((u,\mu)^n\right)$, note that from the spectral decomposition~\eqref{eq:spectral decomposition} we can rewrite
\begin{align}
    \mathbb{P}_n\left((u,\mu)^n\right) = \sum_{s\in \{0,1\}^{n+1}} \frac{1}{2\pi \ii} \oint_{\partial B_\delta(0)} T_{s_1}(z) (\hat{u}+\mu \widehat{w}) \,T_{s_2}(z) ... (\hat{u}+\mu \widehat{w})\,  T_{s_{n+1}}(z) \mathrm{d} z = \sum_{s\in \{0,1\}^{n+1}} T_{s}, \label{eq:rewriting}
\end{align}
where $T_0(z) \coloneqq P/z$, $T_1(z) \coloneqq R^\perp_z(v,\lambda)$, and 
\begin{align}
    T_s \coloneqq \frac{1}{2\pi \ii} \oint_{\partial B_\delta(0)} T_{s_1}(z) (\hat{u}+\mu \widehat{w}) T_{s_2}(z) (\hat{u}+\mu \widehat{w}) ... T_{s_{n}}(z) (\hat{u}+\mu \widehat{w}) T_{s_{n+1}}(z)\mathrm{d}z. \label{eq:T_s def}
\end{align} 
Let us now bound the rank of each $T_s$. First, note that $T_{(1,...,1)} = 0$ because $T_1(z)$ (thus also the integrand
in~\eqref{eq:T_s def}) are holomorphic inside $B_\delta(0)$. On the other hand, if at least one $s_j =0$, then $T_s$ can be written as
\begin{align*}
    T_s = \frac{1}{2\pi \ii} \oint \frac{1}{z^\ell} A(z) P  B(z) \mathrm{d} z
\end{align*}
for holomorphic functions $A(z) \in \mathcal{B}_{1,1}$ and $B(z) \in \mathcal{B}_{-1,-1}$ and $\ell \leq n$. Expanding $A(z)$ as a power series up to order $n-1$, i.e., $A(z) = \sum_{k=0}^{n-1} A_k z^k + z^n A_n(z)$, we find that
\begin{align*}
    T_s = \sum_{k=0}^{n-1} \frac{A_k P }{2\pi \ii} \oint_{\partial B_\delta(0)} \frac{B(z)}{z^{\ell-k}} \mathrm{d} z.
\end{align*}
As $P$ has rank one, each term in the sum has at most rank one, and therefore $\mathrm{rank}(T_s) \leq n$. Using this bound for any element $T_s$ in~\eqref{eq:rewriting}, we conclude that
\begin{align*}
    \mathrm{rank}\left(\mathbb{P}_n\left((u,\mu)^n\right)\right) \leq \sum_{s\in \{0,1\}^{n+1}} \mathrm{rank}(T_s) \leq n 2^{n+1} \leq C^n, 
\end{align*}
for any $C \geq 2 \exp(2/\ee)$, which completes the proof.
\end{proof}

We can now prove the following lemma.
\begin{lemma}[Analytic potential-to-density map] \label{lem:smooth potential to density} The density map $(v,\lambda)\mapsto \rho(v,\lambda)$ belongs to $C^\omega(\mathcal{V}/\{1\}\times \R;\mathcal{D}_N)$. Moreover, its derivative with respect to $v$ at $(v,\lambda)$ is given by
\begin{align}
    D_v \rho =  B_{\Psi(v,\lambda)} R^\perp(v,\lambda) B_{\Psi(v,\lambda)}^\ast + \overline{B_{\Psi(v,\lambda)} R^\perp(v,\lambda) B_{\Psi(v,\lambda)}^\ast} \in \mathcal{B}(\mathcal{V}/\{1\};\mathcal{X}_0), \label{eq:derivative density}
\end{align}
where $B_{\Psi(v,\lambda)}:\mathcal{H}^1_N \rightarrow \mH^1(I;\R)$ is the map defined in~\eqref{eq:Bdef}, $B_{\Psi(v,\lambda)}^\ast : \mathcal{V} \rightarrow \mathcal{H}_N^{-1}$ is the adjoint map in~\eqref{eq:B adjoint def}, and the complex conjugate of an operator is defined as
\begin{align*}
    \overline{B_{\Psi(v,\lambda)} R^\perp(v,\lambda) B_{\Psi(v,\lambda)}^\ast} (f)= \overline{B_{\Psi(v,\lambda)} R^\perp(v,\lambda) B_{\Psi(v,\lambda)}^\ast\bigr(\overline{f}\bigr)},
\end{align*}
where $\overline{f}$ is defined in~\eqref{eq:adjoint potentials}.
\end{lemma}

\begin{proof} By definition, we have
\begin{align*}
    \rho(v,\lambda) = \mathrm{dens}\left(P(v,\lambda)\right),
\end{align*}
where $\mathrm{dens} : \mathcal{S}_1^1 \rightarrow \mH^1(I)$ is the map defined in Section~\ref{sec:density map}.
As $\mathrm{dens}$ is linear, the map $(v,\lambda) \in \mathcal{V}/\{1\}\times \R \rightarrow \rho(v)$ is the composition of an analytic map (by Lemma~\ref{lem:smooth projection}) with a linear one and therefore belongs to $C^\omega(\mathcal{V}/\{1\}\times \R;\mH^1(I;\R))$. Moreover, one can easily check that $\rho$ is normalized as
\begin{align*}
    \int_I \rho(v,\lambda)(x) \mathrm{d} x = \inner{\rho(v,\lambda),1} = \mathrm{Tr} \, P(v,\lambda) \widehat{1} =  \mathrm{Tr}\, P(v,\lambda) N= N,
\end{align*}
and therefore, $(v,\lambda) \mapsto \rho(v,\lambda)$ is analytic from $\mathcal{V}/\{1\}\times \R$ to the affine space $\mathcal{X}_N$. The fact that $\rho(v,\lambda) \in \mathcal{D}_N$ was proved in \cite{Cor26a}. As $\mathcal{D}_N$ is open in $\mathcal{X}_N$, we conclude that $(v,\lambda) \mapsto \rho(v,\lambda) \in C^\omega(\mathcal{V}/\{1\}\times \R; \mathcal{D}_N)$. 

To compute the derivative, let us again simplify the notation by setting $P = P(v,\lambda)$, $R^\perp = R^\perp(v,\lambda)$, and $\Psi = \Psi(v,\lambda)$. Then note that for any $f,g \in \mH^{-1}(I;\C)$ we have
\begin{align*}
    \inner{D_v \rho(f),g} &= \inner{\mathrm{dens}\, D_v P(f),g} \overset{\text{by~\eqref{eq:duality density definition} and~\eqref{eq:derivative formula}}}{=} \mathrm{Tr} \left\{ (R^\perp \hat{f} P + P \hat{f} R^\perp)^\ast \hat{g}\right\} \\
    &= \mathrm{Tr} \, \left\{ P \hat{f}^\ast R^\perp \hat{g} \right\} + \mathrm{Tr} \left\{ R^\perp \hat{f}^\ast P \hat{g}\ \right\} = \inner{\Psi, \hat{f}^\ast R^\perp \hat{g} \Psi} + \inner{\Psi, \hat{g} R^\perp \hat{f}^\ast \Psi}\\
    &\overset{\text{by~\eqref{eq:quantized operators},~\eqref{eq:adjoint potentials}, and~\eqref{eq:B adjoint def}}}{=} \inner{B_{R^\perp \hat{g} \Psi}\Psi, \overline{f}} + \inner{B_{R^\perp \hat{f}^\ast \Psi} \Psi, g} = \overline{\inner{B_\Psi R^\perp \hat{g} \Psi, f}} + \left\langle\overline{B_\Psi R^\perp \hat{f}^\ast \Psi}, g\right\rangle \\
    &\overset{\text{by~\eqref{eq:adjoint resolvent},~\eqref{eq:B adjoint def}, and~\eqref{eq:adjoint potentials}}}{=} \inner{B_\Psi R^\perp B_\Psi^\ast f, g} + \left\langle\overline{B_\Psi R^\perp B_\Psi^\ast \overline{f}}, g\right\rangle, 
\end{align*}
which gives us formula~\eqref{eq:derivative density}. Moreover, since $B_{\Psi}^\ast(1) = N \Psi$, we have $D_v \rho(1) = 2B_{\Psi} R^\perp N \Psi = 0$ so that $D_v \rho$ is a well-defined map in $\mathcal{V}/\{1\}$. Similarly, one can show that $\int_I D_v \rho(u)(x) \mathrm{d} x = \inner{N \Psi,R^\perp B_{\Psi}^\ast u} = 0$, which shows that $D_v \rho$ is indeed a bounded linear operator from $\mathcal{V}/\{1\}$ to $\mathcal{X}_0$. 
\end{proof}

\begin{remark*}[Linear response operator] The operator $D_v \rho$ gives the linear variation of the ground-state density with respect to a variation on the external potential. Hence, we shall call it the linear response operator (LRO). 
\end{remark*}

\begin{remark*}[Simplified formula in the real-valued case] The LRO as defined in~\eqref{eq:derivative density} can be seen as a $\C$-linear operator from $\mH^{-1}(I)/\{1\}$ to $\mathcal{X}_0^\C$. However, note that, when $v$ and $w$ are real-valued, the Hamiltonian $H_N(v,w)$ is real. In particular, the ground-state $\Psi$ has a global phase and the reduced resolvent commutes with complex conjugation. Consequently, formula~\eqref{eq:derivative density} simplifies to
\begin{align}
    D_v\rho(f) = 2 B_\Psi R^\perp B_\Psi^\ast f, \quad \mbox{for $f\in \mathcal{V}/\{1\}$.} \label{eq:simplified LRO}
\end{align}
\end{remark*}

We now prove that the LRO is invertible.

\begin{lemma}[Inverse linear response]\label{lem:inverse derivative} For any $(v,\lambda)\in \mathcal{V}/\{1\}\times \R$, the linear response operator $D_v \rho \in \mathcal{B}(\mathcal{V}/\{1\};\mathcal{X}_0)$ is invertible. 
\end{lemma}

\begin{proof} First, we note that, since the map $\rho \mapsto v(\rho;\lambda)$ is Lipschitz in $\rho$ by the QHK Theorem~\ref{thm:QHK}, we have 
\begin{align*}
    \norm{\epsilon D_v \rho(u)}_{\mH^1} = \norm{\rho(v+\epsilon u) - \rho(v) + o(\epsilon)}_{\mH^1} \gtrsim \epsilon \norm{u}_{\mathcal{V}/\{1\}} + o(\epsilon),
\end{align*}
where $o(\epsilon)$ denotes a remainder $R(\epsilon)$ satisfying $\lim_{\epsilon \searrow 0} R(\epsilon)/\epsilon =0$. Hence, dividing by $\epsilon$ and taking the limit $\epsilon \searrow 0$, we conclude that
\begin{align}
    \norm{D_v \rho(u)}_{\mH^1} \gtrsim \norm{u}_{\mathcal{V}/\{1\}}, \quad \mbox{for any $u \in \mathcal{V}/\{1\}$.} \label{eq:lower bound est}
\end{align}
Thus $D_v\rho$ is injective and has closed range in $\mathcal{X}_0$. Hence, to complete the proof, it suffices to show that $D_v \rho$ has dense range in $\mathcal{X}_0$. For this, let $u \in \mathcal{V}/\{1\}$ be such that 
\begin{align*}
    \inner{D_v \rho(g),u} = 0 \quad \mbox{for any $g \in \mathcal{V}/\{1\}$,} 
\end{align*}
where $\inner{\cdot,\cdot}$ is the dual pairing in $\mH^1 - \mH^{-1}$ and we use the isometric
identification
\begin{align}
  \mathcal{X}_0^*\simeq \mH^{-1}(I;\R)/\mathrm{span}\{1\}=\mathcal{V}/\{1\},
  \qquad [u]\longmapsto\bigl(f \in \mathcal{X}_0 \mapsto\inner{f,u}\bigr).
\end{align}
Then, it suffices to show that $u=0$. For this, note that by the simplified formula~\eqref{eq:simplified LRO}, equation~\eqref{eq:B adjoint def}, and estimate~\eqref{eq:strictly negative}, we have
\begin{align*}
    0 = -\inner{D_v \rho(u),u} = -2\inner{R^\perp B_{\Psi(v,\lambda)}^\ast u, B_{\Psi(v,\lambda)}^\ast u} \gtrsim \norm{P_\Psi^\perp B_\Psi^\ast u}_{\mathcal{H}_N^{-1}}^2,
\end{align*}
and therefore $P_\Psi^\perp B_\Psi^\ast u = 0$. Hence $\inner{P_\Psi^\perp B_\Psi u, \Phi} = \inner{u, B_\Psi^\ast \Phi}$ for any $\Phi \in \mathcal{H}_N^1\cap \{\Psi\}^\perp$. Since $B_\Psi^\ast:\mathcal{H}_N^1 \cap \{\Psi\}^\perp \rightarrow \mathcal{X}_0$ is surjective by Lemma~\ref{lem:inverse}, this implies that $u=0$, and therefore the range of $D_v\rho$ is dense. This completes the proof.
\end{proof}

We can now complete the proof of Theorem~\ref{thm:analytic map}.
\begin{proof}[Proof of Theorem~\ref{thm:analytic map}] Fix $(\rho_0,\lambda_0)$ and let $U \subset \mathcal{X}_0$ be an open neighborhood of $0$ such that $\rho_0 + U \subset \mathcal{D}_N$ and $v_0 = v(\rho_0,\lambda_0)$. Define $G: (U \times \R) \times \mathcal{V}/\{1\} \rightarrow \mathcal{X}_0$ as
\begin{align*}
    G\left((\sigma,\lambda),v\right) = \rho(v+v_0,\lambda+\lambda_0) - \rho_0-\sigma,
\end{align*}
where $\rho(v,\lambda)$ is the potential-to-density map. By Lemma~\ref{lem:smooth potential to density}, the map $G$ is analytic and satisfies $G((0,0),0) = 0$. Moreover, by Lemma~\ref{lem:inverse derivative}, the Fr\'echet derivative 
\begin{align*}
    D_v G\left((0,0),0\right) = D_v\rho (v_0,\lambda_0) : \mathcal{V}/\{1\} \rightarrow \mathcal{X}_0
\end{align*}
is invertible. We can therefore apply the implicit function theorem (cf. Theorem~\ref{thm:IFT}) to conclude that there exists a unique analytic map $\tilde{v}: V \times J \rightarrow \mathcal{V}/\{1\}$ on a neighborhood $(0,0) \in V \times J \subset U \times \R$ such that $G\left((\sigma,\lambda), \tilde{v}(\sigma,\lambda)\right) = 0$. By the uniqueness of the representing potential (cf. HK Theorem~\ref{thm:QHK}), it follows that $v(\rho_0 + \sigma,\lambda_0 +\lambda) = \tilde{v}(\sigma,\lambda) + v_0$ and therefore $v$ is analytic around $(\rho_0,\lambda_0)$. As $(\rho_0,\lambda_0) \in \mathcal{D}_N \times \R$ is arbitrary, we have $v \in C^\omega(\mathcal{D}_N \times \R;\mathcal{V}/\{1\})$, thereby completing the proof.
\end{proof}

\section{Proof of corollaries}
\label{sec:corollaries proof}

We now prove Corollaries~\ref{cor:analytic levy-lieb},~\ref{cor:exchange} and~\ref{cor:GL perturbation series}, and Theorem~\ref{thm:holomorphic DFT}.

\begin{proof}[Proof of Corollary~\ref{cor:analytic levy-lieb}] As $v(\rho;\lambda)$ is analytic, the map $(\rho,\lambda) \mapsto H_N(v(\rho;\lambda),\lambda w) \in \mathcal{B}_{1,-1}$ is analytic since it is a composition of linear and analytic maps. Similarly, the map sending $(\rho,\lambda)$ to the ground-state projector $P(v(\rho;\lambda),\lambda) \in \mathcal{S}_1^1$ is analytic, as it is a composition of analytic maps (see Lemma~\ref{lem:smooth projection}). Thus
\begin{align}
F_{\rm LL}(\rho;\lambda)
  =\Tr\!\left[
      P\bigl(v(\rho;\lambda),\lambda\bigr)
      H_N(0,\lambda w)
    \right]
\end{align}
is analytic.
\end{proof}

\begin{proof}[Proof of Corollary~\ref{cor:exchange}] Recalling that $T_{\rm KS}(\rho) = F_{\rm LL}(\rho;0)$ for $\rho \in \mathcal{D}_N$ (see~\cite[Lemma 6.1]{Cor26a}), we see that 
\begin{align*}
    E_{\rm xc}(\rho;\lambda) = \frac{F_{\rm LL}(\rho;\lambda) - F_{\rm LL}(\rho;0)}{\lambda} - E_H(\rho),
\end{align*}
and therefore, $E_{\rm xc}$ is analytic. Consequently, $E_{\rm x}(\rho) = E_{\rm xc}(\rho;0)$ is well-defined and analytic in $\rho$. Identity~\eqref{eq:exchange formula} simply follows from the fact that the derivative order does not matter, i.e., $\partial_\lambda D_\rho F_{\rm LL} = D_\rho \partial_\lambda F_{\rm LL}$.
\end{proof}

\begin{proof}[Proof of Corollary~\ref{cor:GL perturbation series}] As the correlation energy $E_c(\rho;\lambda) = E_{\rm xc}(\rho;\lambda) - E_x(\rho)$ is analytic in $(\rho, \lambda)$ by Corollaries~\ref{cor:analytic levy-lieb} and~\ref{cor:exchange}, it admits an absolutely convergent series expansion in $\lambda$ around any $\lambda\in \R$. Moreover, using $F(\lambda)=F_{\rm LL}(\rho;\lambda)$, the definitions of $E_{\rm xc}, E_{\rm x}$ and $E_{\rm H}$ give the exact identity
\begin{align*}
 \lambda E_{\rm c}(\rho;\lambda) \coloneqq \lambda E_{\rm xc}(\rho;\lambda) - \lambda E_{\rm x}(\rho;w)
 =F(\lambda)-F(0)-\lambda F'(0).
\end{align*}
It then follows that 
\begin{align}
 \lambda E_{\rm c}(\rho;\lambda)
 =\frac{\lambda^2}{2}F''(0)+O(\lambda^3), \label{eq:second order exp}
\end{align}
so $E_{\rm c}=O(\lambda)$. In particular, taking derivative with respect to $\rho$ implies that $v_c(\rho;\lambda) = \mathcal{O}(\lambda)$ as well.

To compute the explicit formula for $E_{
\rm c}^{
\rm GL2}$, let $\Psi_0$ be the ground state of the Kohn-Sham Hamiltonian $H_{\rm KS} \coloneqq H_N(v(\rho;0),0)$, $\{\Psi_j\}_{j\geq1}$ its excited eigenstates, and $\omega_j\coloneqq E_j-E_0$ the excitation energies. Then note that $\partial_\lambda v(\rho;0)=-v_{\mathrm H}(\rho)-v_{\rm x}(\rho)$ by~\eqref{eq:exchange formula}. Hence, the first-order perturbation (in $\lambda$) of the density-fixed Hamiltonian $H_N(v(\rho;\lambda), \lambda w)$ is
\begin{align*}
 H_N(v(\rho;\lambda),\lambda w) = H_{\rm KS} + \lambda W_1 + \mathcal{O}(\lambda^2), \quad W_1 \coloneqq \widehat w+\widehat{\partial_\lambda v}(\rho;0)
 =\widehat w-\widehat v_{\rm x}(\rho)-\widehat v_{\mathrm H}(\rho).
\end{align*}
Second-order analytic perturbation theory for the simple eigenvalue gives the
expansion
\begin{align*}
 E(\lambda) \coloneqq E(v(\rho;\lambda),\lambda w) = \inner{\Psi_0, H_N(v(\rho;\lambda),\lambda w) \Psi_0}-\lambda^2 \sum_{j\geq1}\frac{|\inner{\Psi_0,W_1\Psi_j}|^2}{\omega_j} + \mathcal{O}(\lambda^3)
\end{align*}
for the ground-state energy of $H_N(v(\rho;\lambda),\lambda w)$. As $\rho_{\Psi_0} =\rho$, we find that
\begin{align*}
F_{\rm LL}(\rho;\lambda)
= E(\lambda)-\inner{\rho,v(\rho;\lambda)} = \inner{\nabla \Psi_0, \nabla\Psi_0} + \lambda \inner{\Psi_0, \hat{w} \Psi_0} - \lambda^2 \sum_{j\geq 1} \frac{|\inner{\Psi_0, W_1 \Psi_j}|^2}{\omega_j} + \mathcal{O}(\lambda^3).
\end{align*}
We thus have
\begin{align*}
\frac12F''(0)
 =-\sum_{j\geq1}
   \frac{\left|\inner{\Psi_0,
     (\widehat w-\widehat v_{\rm x}-\widehat v_{\mathrm H})\Psi_j}
   \right|^2}{\omega_j},
\end{align*}
which together with~\eqref{eq:second order exp} yields the asserted formula for $E_c^{\rm GL2}$.
\end{proof}

We now turn to the proof of Theorem~\ref{thm:holomorphic DFT}. 

\begin{proof}[Proof of Theorem~\ref{thm:holomorphic DFT}] First, from Proposition~\ref{prop:holomorphic extension}, there exists a holomorphic extension of the Levy--Lieb functional $F_\C$ to a suitable complex neighborhood $\widetilde{U} \subset \mathcal{X}_N^\mathbb{C} \times \C$ of $\mathcal{D}_N \times \R$. After possibly shrinking $\widetilde{U}$, we can assume this set is connected. The uniqueness of $F_\C$ follows from the corresponding uniqueness in Proposition~\ref{prop:holomorphic extension}. We have thus established properties~\ref{it:extension} and~\ref{it:uniqueness}. 

To prove properties~\ref{it:bijective potential map} and~\ref{it:Hamiltonian relation}, we first note that, by applying Proposition~\ref{prop:holomorphic extension}, there exists a holomorphic extension $P_\C$ of the ground-state projection map~\eqref{eq:projection-valued map} to an open subset $\widetilde{V} \subset \mH^{-1}/\{1\} \times \C$ containing $\mathcal{V}/\{1\} \times \R$. Now define $G: \widetilde{U} \rightarrow \mH^{-1}/\{1\} \times \C$ and $L : \widetilde{V} \rightarrow \mathcal{X}_N^\mathbb{C} \times \C$  as
\begin{align*}
    G(\rho,\lambda) \coloneqq (- D_\rho F_\C(\rho;\lambda), \lambda) \quad \mbox{and}\quad L(v,\lambda) \coloneqq (\mathrm{dens} \, P_\C(v,\lambda), \lambda).
\end{align*}
Next, note that $\mathcal{V}/\{1\} \times \R\subset L^{-1}(\mathcal{D}_N\times \R )$ because the map $v \mapsto \mathrm{dens}\, P_\C(v,\lambda)$ maps $\mathcal{V}/\{1\}$ bijectively to $\mathcal{D}_N$ for each $\lambda \in \R$. We can thus define $V$ as the open connected component of $\widetilde{V} \cap L^{-1} (\widetilde{U})$ containing $\mathcal{V}/\{1\} \times \R$ and consider only the restriction $L: V \mapsto \widetilde{U}$. By construction, the composition $G\circ L: V \rightarrow \mH^{-1}/\{1\} \times \C$ is well-defined. Moreover, we have
\begin{align}
    G \circ L(v,\lambda) = \left(- D_\rho F_{\rm LL}\left(\mathrm{dens}\, P(v,\lambda), \lambda\right), \lambda\right) = (v,\lambda) \quad \mbox{for any $(v,\lambda) \in \mathcal{V}/\{1\} \times \R$.} \label{eq:identity on real}
\end{align}
Since $V$ is open and connected, from the identity principle (cf. Proposition~\ref{prop:identity theorem}), the identity~\eqref{eq:identity on real} holds everywhere in $V$. We can now define $U$ as the open connected component of $G^{-1}(V)$ containing $\mathcal{D}_N \times \R$. Note that, since $L(V)$ is connected (as $L$ is continuous), contains $\mathcal{D}_N \times \R$, and satisfies $L(V) \subset G^{-1}(V)$ (since~\eqref{eq:identity on real} holds in $V$), we must have $L(V) \subset U$. Thus $G: U \rightarrow V$ is surjective. On the other hand, we can again apply the identity principle to conclude that $L \circ G(\rho,\lambda) = (\rho,\lambda)$ for any $(\rho,\lambda) \in U$. Thus $G : U \rightarrow V$ is in fact a holomorphic bijection with holomorphic inverse $G^{-1} =L$. This proves~\ref{it:bijective potential map}. 

The remaining statements in~\ref{it:Hamiltonian relation} now follow from the identity principle. More precisely, as
\begin{align*}
    P_\C(v,\lambda) P_\C(v,\lambda) - P_\C(v,\lambda) = 0 \quad \mbox{and} \quad \mathrm{Tr} \, P_\C(v,\lambda) = 1, \quad \mbox{for any $(v,\lambda) \in \mathcal{V}/\{1\} \times \R$,}
\end{align*}
and $V$ is connected, these equations hold for all $(v,\lambda) \in V$ by Proposition~\ref{prop:identity theorem}. Hence, recalling that any (norm) continuous projection-valued map has locally constant rank, we conclude that $P_\C(v,\lambda)$ is the desired rank-one projection-valued map in~\ref{it:Hamiltonian relation}. Similarly, equation~\eqref{eq:eigenvalue property} follows from the identity principle.
\end{proof}

\section{Conclusion}\label{sec:conclusion}

In this paper, we established a quantitative version of the Hohenberg--Kohn theorem for spinless fermions living in a compact one-dimensional interval subjected to external potentials in a specific class of distributions. Applying this result, we then showed that the universal (constrained-search) functional and the density-to-potential map are analytic with respect to both the density, endowed with the natural Sobolev topology, and the interaction strength. As applications, we derived the existence of an exchange-only potential and justified the G\"orling--Levy perturbation series in this simplified setting.

As a remarkable consequence of these results, we showed that the DFT framework can be extended to the complex domain. More precisely, we showed that the density-to-potential map can be extended to a bi-holomorphic map between suitable subspaces of complex-valued potentials and complex-valued densities. This seems to be the first result in this direction in the literature. In particular, it raises several questions that we could not address here.
\begin{enumerate}
    \item What is the (physical) interpretation of the eigenvalue associated with the holomorphic spectral projector in Theorem~\ref{thm:holomorphic DFT}~\ref{it:Hamiltonian relation}? For instance, for complex potentials that are small perturbations of real-valued ones, it is clear that such eigenvalues correspond to the eigenvalues with lowest real part. In particular, the associated spectral projector corresponds to the dominating mode in the large time asymptotics of the heat semigroup. However, it is not clear whether this is the case for all potentials in the domain of $F_\C$.
    \item Can one characterize the maximal extension domains of $F_{\rm LL}$? A similar question is, can one solve the $\mathcal{V}$-representability problem in the complex-valued case, i.e., characterize the set of all ground-state densities of Hamiltonians with complex-valued potentials and fixed complex interaction? To properly pose this question, one first needs to clarify the notion of ground-state in the complex case.
    \item How can one leverage the analyticity of $F_{\rm LL}$ to construct new (and justify current) xc-approximations? Here, we made a first small step in this direction by justifying the GL perturbation formula. In this regard, a natural follow-up question is whether one can estimate the radius of convergence of this series, which is intimately connected with the previous question of characterizing the maximal extension domains of $F_{\C}$.
\end{enumerate}

Another important consequence of our results is that the inverse problem of retrieving the external potential that generates a target density is not only well-posed in the Hadamard sense but also Lipschitz stable, for both the \emph{interacting} and non-interacting cases. This result also raises some interesting questions:
\begin{enumerate}
    \item Can one give explicit estimates for the Lipschitz constant in Theorem~\ref{thm:QHK}? Note that our proof relies on a compactness argument (see Lemma~\ref{lem:inverse}), and does not provide an estimate on this constant. Such estimates could be useful to obtain error estimates in numerical implementations of the inverse scheme.
    \item What is the connection of the QHK presented here with the Moreau--Yosida-regularization of DFT studied in \cite{PL26}? More precisely, the result in \cite[Corollary 6.3]{PL26} could be interpreted as a regularized version of the QHK presented here. However, in the regularized setting, the choice of topology is rather flexible.
    In particular, it is not clear how the constant in \cite[Corollary 6.3]{PL26}, for a chosen topology, relates to the Lipschitz constant for the unregularized setting in the limit $\varepsilon \downarrow 0$.
\end{enumerate}
However, we emphasize that the results presented here are restricted to the one-dimensional setting with Neumann boundary conditions, and we do \emph{not} know how (or if it is possible) to extend them to higher dimensions. In fact, even an extension to the one-dimensional case with Dirichlet boundary conditions seems to be non-trivial.

Finally, let us conclude by mentioning that the results obtained here should also have implications for the time-dependent variant of DFT (TDDFT) \cite{RG84,MMN+12}. More precisely, the invertibility of the linear response operator (cf. Lemma~\ref{lem:inverse derivative}) can likely be extended to the dynamic (or time-dependent) linear response operator in the frequency domain, which appears in the Dyson equation of linear response TDDFT \cite{Cor24,CDF25,DLL25}. This result could potentially be used to establish a rigorous foundation for TDDFT in the linear response regime, at least in the one-dimensional setting, for which no consensus on a (formal) proof seems to exist to this date \cite{vLe99,MMN+12,FLL+16,Sch25}. However, investigating this question would go far beyond the scope of this paper, and we reserve it for future contributions.

\addtocontents{toc}{\protect\setcounter{tocdepth}{-1}}
\section*{Acknowledgements}
The authors thank Mihály Andras Csirik and Erik Ingemar Tellgren for fruitful discussions on the content of the article.

T.C.~Corso acknowledges funding by the \emph{Deutsche Forschungsgemeinschaft} (DFG, German Research Foundation) - Project number 442047500 through the Collaborative Research Center ``Sparsity and Singular Structures'' (SFB 1481) and Project number 572811220. AL received funding from the ERC-2021-STG under grant agreement No.~101041487 REGAL. AL was also supported by the Research Council of Norway through CoE Hylleraas Centre for Quantum Molecular Sciences Grant No.~262695.

\addtocontents{toc}{\protect\setcounter{tocdepth}{2}}
\appendix
\section{Analytic functions in Banach spaces}
\label{app:analytic}
In this section, we recall the definition and some elementary properties of analytic functions between Banach spaces. We then state and briefly sketch the proof of the implicit function theorem for such functions. These results are well-known and can be found in standard references, e.g.,  \cite{Muj86,Din99,KP13}.

\begin{definition}[Analytic functions] \label{def:analytic}
Let $X, Y$ be real Banach spaces and $F: U \rightarrow Y$ a function defined on an open set $U\subset X$. Then we say $F$ is analytic at $x_0\in U$ if there exists a sequence of $\R$-multi-linear operators $F_{x_0}^k:X^k 
\rightarrow Y$ and constants $C,R>0$ such that
\begin{align}
    \norm{F_{x_0}^k}_{X^k \rightarrow Y} = \sup_{x_1,...,x_k \in X\setminus\{0\}} \frac{\norm{F_{x_0}^k(x_1,...,x_k)}_Y}{\norm{x_1}_X ... \norm{x_k}_X} \leq  CR^k, \quad \mbox{for any $k\geq 0$}, \label{eq:analytic est}
\end{align}
and
\begin{align}
    F(x) = \sum_{k\geq 0} F^k_{x_0}(x-x_0,...,x-x_0), \quad \mbox{for any $x\in B_{1/R}(x_0) \cap V$ for some neighborhood $x_0 \in V \subset U$.} \label{eq:power series}
\end{align}
We say that $F$ is analytic in $U$ if it is analytic at every point in $U$. Moreover, we denote the space of analytic functions from $U$ to $Y$ by $C^\omega(U;Y)$.
\end{definition}

 Throughout the paper, we shall mostly deal with functions defined on a relatively open subset of an affine closed subspace of a Banach space $X$. In this case, the definition of analytic function should be understood as follows.
 
 \begin{definition}[Analytic functions on affine spaces] Let $F: U \rightarrow Y$ be a function from a relatively open set $U\subset V$, where $V$ is an affine closed subspace of a Banach space $X$. Then we say that $F$ is analytic around $x_0 \in U$ if the shifted function $\tau_{x_0}F : U-x_0 \rightarrow Y$, $(\tau_{x_0}F)(x) = F(x+x_0)$, is analytic around $0$ as a function from the Banach space $V-x_0$ to $Y$. If $F$ is analytic around any point $x_0 \in U$, then we say that $F$ is analytic in $U$ and belongs to $C^\omega(U;Y)$.
 \end{definition}

\begin{remark*}[Symmetric multi-linear operators]
    Since only terms of the form $(x-x_0,...,x-x_0) \in X^n$ are used as arguments in the Taylor expansion of $F$, we can always assume the multi-linear operators $F^n_{x_0}$ to be symmetric, i.e., $F^n_{x_0}(x_1,...,x_N) = F^n_{x_0}(x_{\sigma(1)}, ..., x_{\sigma(N)})$ for any permutation $\sigma \in \mathcal{S}_N$. Moreover, with this convention, the Taylor series uniquely determines each $F_{x_0}^n$ via the polarization formula~\eqref{eq:polarization formula} stated below.
\end{remark*}

\subsection{Complexification and the identity theorem} For a real Banach space $X$, we define its complexification, $X_\C$, as the complex Banach space
\begin{align*}
    X_\C \coloneqq X\times X, \quad\mbox{endowed with the norm} \quad 
    \norm{(x,y)}_{X_\C} \coloneqq \sup_{\theta\in[0,2\pi]} \norm{x\cos\theta-y\sin\theta}_X
\end{align*}
and the scalar multiplication by a complex number
\begin{align*}
    \lambda (x, y) = \left((\alpha x- \beta y), (\beta x + \alpha y)\right), \quad \mbox{for $\lambda = \alpha + \ii \beta \in \C$.}
\end{align*}
Following standard practice, we use the notation $z=x+\ii y \in X_\C$ for an element $z = (x,y) \in X_\C$. 

On complex Banach spaces, the definition of analytic functions can be extended by replacing the $\R$ linear functionals by $\C$ linear ones in~\eqref{eq:power series}. 

\begin{definition}[Holomorphic functions] Let $X_\C$ and $Y_\C$ be two complex Banach spaces, and $U\subset X_\C$ an open subset. Then we say that a function $F : U \subset X_\C \rightarrow Y_\C$ is holomorphic on the open set $U\subset X_\C$ if and only if, around any point $x_0 \in U$, $F$ can be locally written as a power series as in~\eqref{eq:power series} with symmetric $\C$-multi-linear operators satisfying estimate~\eqref{eq:analytic est}.
\end{definition}

We now recall two algebraic facts. First, for any symmetric multi-linear functional, we have the polarization formula (cf. \cite[Theorem 1.10]{Muj86} or \cite[Corollary 1.6]{Din99})
\begin{align}
    F(x_1,...,x_n) = \frac{1}{n! 2^n}  \sum_{s\in \{-1,1\}^n} s_1...s_n F\left(\sum_{j=1}^n s_j x_j,\sum_{j=1}^n s_j x_j,...,  \sum_{j=1}^n s_j x_j\right). \label{eq:polarization formula}
\end{align}
Second, for any $\C$-multi-linear functional $F:X_\C\times ... \times X_\C \rightarrow Y_\C$ we have
\begin{align}
    F(x_1^0 + \ii x_1^1, x_2^0 + \ii x_2^1,...,x_n^0 + \ii x_n^1) = \sum_{s\in \{0,1\}^n} \ii^{\sum s_j} F(x_1^{s_1},x_2^{s_2},...,x_n^{s_n}), \quad \mbox{for any $x_j^0,x_j^1 \in X$.} \label{eq:multi-linear real expansion}
\end{align}

Using these facts, we can prove the following simple Banach space variant of the well-known identity principle in complex analysis.

\begin{proposition}[Identity principle on real sets] \label{prop:identity theorem} Let $U \subset X_\C$ be a connected open set and $F,G:U\rightarrow Y_\C$ be holomorphic functions such that $F = G$ on a subset $U' \subset U \cap X$ with non-empty interior in $X$. Then $F(x) = G(x)$ for every $x\in U$. 
\end{proposition}

\begin{proof} It suffices to show that, if $F = 0$ on some $U'$ as in the proposition, then $F=0$ everywhere in $U$. For this, let $x_0 \in U'$ and note that
\begin{align*}
    F^n_{x_0}(x,...,x) = \frac{1}{n!} \frac{\mathrm{d}^n}{\mathrm{d} t^n} F(x_0+tx)\bigr\rvert_{t=0} = 0\quad \mbox{for any $x\in B_\delta(0) \subset X$,}
\end{align*}
where $F^n_{x_0}$ are the $\C$-linear operators in~\eqref{eq:power series}. As $F^n_{x_0}$ is symmetric, by the polarization formula~\eqref{eq:polarization formula} we have $F^n_{x_0}(x_1,...,x_n) = 0$ for any $x_j \in X$. In particular $F^n_{x_0}(x_1,...,x_N) = 0$ for any $x_j \in X_\C$ by~\eqref{eq:multi-linear real expansion}. Hence $F^n_{x_0} = 0$ for any $n\in \N$, and therefore, by~\eqref{eq:power series}, $F = 0$ on a complex neighborhood $B_\delta^\C(x_0) \subset U$. Hence, the set $V \coloneqq \{ x\in U : \mbox{$F$ has identically zero Taylor series around $x$} \}$ is open. Since $U$ is connected and $V\neq \emptyset$, to complete the proof it suffices to show that $U\setminus V$ is open as well. 

For this, note that $U\setminus V$ are the points with Taylor series that are not identically vanishing. Hence, for any $x\in U\setminus V$, there exists $m \in \N$ such that $F^m_{x} \neq 0$. Since the Taylor series is uniquely defined at each point, by regrouping the terms in~\eqref{eq:power series} and using the symmetry we find that
\begin{align*}
    F^m_{y}(z,...,z) = \sum_{k\geq m} \binom{k}{m} F^k_x(\overbrace{z,...,z}^{m \text{ times}},y-x,...,y-x) = F_x^m(z,...,z) + \sum_{k\geq m+1} \binom{k}{m} F^k_x(\overbrace{z,...,z}^{m \text{ times}},y-x,...,y-x).
\end{align*}
Hence, using the bounds in~\eqref{eq:analytic est}, it follows that
\begin{align*}
    \norm{F^m_x - F^m_y} \lesssim  R\norm{y-x}, \quad \mbox{for $\norm{y-x}$ small enough.}
\end{align*}
Therefore, the Taylor series around $y$ does not vanish identically, for any $y$ close enough to $x$. This shows that $U\setminus V$ is also open, which completes the proof.
\end{proof}

As a consequence, any analytic function can be extended to a holomorphic function on a sufficiently small complex neighborhood of the original domain.
\begin{proposition}[Holomorphic extension] \label{prop:holomorphic extension} Let $U\subset X$ be an open subset of the real Banach space $X$ and $F:U  \rightarrow Y$ analytic. Then there exists an open set $U_\C \subset X_\C$ and a holomorphic function $F_\C : U_\C \rightarrow Y_\C$ such that $U \subset U_\C$ and
\begin{align*}
    F_\C(x) = F(x), \quad \mbox{for any $x\in U$.}
\end{align*}
Moreover, the extension is unique in the sense that, if $F_\C':U_\C' \rightarrow Y_\C$ is another holomorphic extension, then $F_\C = F_\C'$ on the connected component of $U_\C \cap U_\C'$ containing $U$.
\end{proposition}

\begin{proof} For any $x_0\in U$ and $k \in \N$, we can extend $F^k_{x_0}$ to a $\C$-linear functional $F_{\C,x_0}^k$ via formula~\eqref{eq:multi-linear real expansion}. It is easy to verify that $\norm{F_{\C,x_0}^k} \leq 2^k \norm{F^k_{x_0}}$. Hence, the extended function
\begin{align*}
    F_{\C}^{x_0} = \sum_{k\geq 0} F_{\C,x_0}^k(x-x_0,...,x-x_0),
\end{align*}
defines a holomorphic function on the complex neighborhood $B_{1/R_{x_0}}(x) \coloneqq \{x \in X_\C : \norm{x-x_0}_{X_\C} < 1/R_{x_0}\} \subset X_\C$, where $R_{x_0}$ is the radius of convergence of the series around $x_0$. Since this extension is independent of the base point $x_0$, i.e., $F_{\C}^{x_0}(x) = F_{\C}^{x_1}(x)$ on $B_{1/{R_{x_0}}}(x_0) \cap B_{1/R_{x_1}}(x_1)$ (by the identity principle in Proposition~\ref{prop:identity theorem}), we obtain a holomorphic extension $F_\C$ to the neighborhood $U_\C \coloneqq \cup_{x_0 \in U} B_{1/R_{x_0}}(x_0)$. The uniqueness statement follows from Proposition~\ref{prop:identity theorem}.
\end{proof}

\subsection{Implicit function theorem} The following version of the implicit function theorem
for analytic functions on real Banach spaces will be useful here. For the sake of completeness, we briefly sketch the proof below. 

\begin{theorem}[Implicit Function Theorem -  Analytic version]\label{thm:IFT} Let $X,Y,Z$ be real Banach spaces, $U \subset X$ and $V\subset Y$ be open subsets, and $F: U\times V \subset X \times Y \rightarrow Z$ be a function such that $F \in C^{\omega}(U\times V;Z)$, $F(0,0) = 0$, and 
\begin{align*}
    D_Y F(0,0) \in \mathcal{B}(Y,Z) \quad \mbox{has a continuous inverse.}
\end{align*}
Then there exists a ball centered at the origin $B_\delta \subset X$ and a unique continuous function $f:B_\delta \rightarrow Y$ such that
\begin{align*}
   f(0) = 0 \quad\mbox{and}\quad  F\left(x,f(x)\right) = 0, \quad \mbox{for any $x\in B_\delta$.}
\end{align*}
Moreover, $f \in C^{\omega}(B_\delta;Y)$. 
\end{theorem}

\begin{proof}[Proof sketch of Theorem~\ref{thm:IFT}] Let $F_\C$ be a holomorphic extension to an open set $(0,0)\in U_\C \times V_\C \subset X_\C \times Y_\C$, which exists by Proposition~\ref{prop:holomorphic extension}. Now note that the partial derivative of $F_\C$ with respect to $Y_\C$ at $(0,0)$ can be represented as
\begin{align*}
    D_0 F_\C(0,0) = \begin{pmatrix} D_0 F(0,0) & 0 \\ 0 & D_0 F(0,0)
    \end{pmatrix}
\end{align*}
in the block decomposition with respect to $X_\C = X \oplus X$. Thus since $D_0 F(0,0)$ is invertible in $\mathcal{B}(Y,Z)$, it follows that $D_0 F_\C(0,0) \in \mathcal{B}(Y_\C;Z_\C)$ is also invertible in the complexified spaces. Hence, by the standard implicit function theorem on Banach spaces (cf.\cite{KP13}[Theorem 3.4.10]), there exists an open ball $B_\delta^\C \subset X_\C$ and a unique Fr\'echet differentiable function 
\begin{align*}
    f_\C:B_\delta^\C(0) \rightarrow Y_\C \quad \mbox{such that $f_\C(0) = 0$ and $F_\C(x,f_\C(x)) = 0$ for any $x\in B_\delta^\C(0)$.}
\end{align*}
 Similarly, applying the same theorem to $F$, we find a unique continuous $f$ defined on $B_\delta(0) \subset X$ satisfying $F(x,f(x)) = 0$. As $y = f^\C(x)$ is the unique solution of $F_\C(x,y)$ for any $x\in B^\C_\delta(0)$ close enough to $0$ (which follows from a contraction fixed point argument), we have $f_\C(x) = f(x)$ for any $x\in X \cap B^\C_\delta(0) =  B_\delta(0)$. In particular, $f$ is the restriction to $B_\delta$ of a Fr\'echet differentiable function on an open set of the complex Banach space $X_\C$. It then follows from \cite{Muj86}[Theorem 13.16] that $f_\C$ is holomorphic, and therefore, $f$ is analytic.
\end{proof}

\section*{Data availability}
No datasets were generated or analyzed during the current study.

\section*{Competing interests}

The authors have no competing interests to declare that are relevant to the content of this article.


\bigskip
\end{document}